%% file: main.tex
\documentclass[twocolumn]{article}

\input{common}

\title{A Liquid Perspective on Democratic Choice \\ ~\\
	{\large\em preliminary work-in-progress; may become
		\href{https://bford.info/book/}{part of a future book} \\
	written and first distributed November 2018 
	in discussions for the edited volume \\
	\href{https://press.uchicago.edu/ucp/books/book/chicago/D/bo68657177.html}{Digital Technology and Democratic Theory}
	\\ ~}}
\author{Bryan Ford}
\date{}

\begin{document}
\maketitle

\input{liquid/abs}

\tableofcontents

\input{liquid/top}

\arxiv{
\bibliography{soc,sec,theory}
}{
\bibliography{main}
}
\bibliographystyle{plain}

\end{document}

%% file: common.tex
\usepackage{fullpage}
\usepackage{times}
\usepackage{cite} 
\usepackage[breaklinks,colorlinks,linkcolor=blue,citecolor=blue,urlcolor=blue]{hyperref}
\usepackage[hyphenbreaks]{breakurl}
\usepackage{xspace}
\usepackage{color}
\usepackage{graphicx}
\usepackage{epigraph}
\usepackage{subfig}
\usepackage{breakcites}

\long\def\com#1{}

\long\def\xxx#1{}

\long\def\arxiv#1#2{#2}		

\newcommand{\ie}{{\em i.e.}\xspace}
\newcommand{\eg}{{\em e.g.}\xspace}

%% file: liquid/abs.tex
\begin{abstract}
The idea of liquid democracy responds to a widely-felt desire
to make democracy more ``fluid'' and continuously participatory.
Its central premise is to enable users to employ networked technologies
to control and {\em delegate} voting power,
to approximate the ideal of direct democracy
in a scalable fashion that accounts for time and attention limits.
There are many potential definitions, meanings, and ways to implement
liquid democracy, however, and many distinct purposes
to which it might be deployed.
This paper develops and explores the ``liquid'' notion and what it might mean
for purposes of enhancing voter choice by spreading voting power,
improving proportional representation systems,
simplifying or aiding voters in their choice,
or scaling direct democracy through specialization.
The goal of this paper is to disentangle and further develop
some of the many concepts and goals that liquid democracy ideas often embody,
to explore their justification with respect to existing democratic traditions
such as transferable voting and political parties,
and to explore potential risks in liquid democracy systems
and ways to address them.
\end{abstract}

%% file: liquid/top.tex
\input{liquid/intro}

\input{liquid/spread}

\input{liquid/trans}

\input{liquid/deleg}

\input{liquid/spec}

\input{liquid/time}

\input{liquid/impl}

\xxx{
 \subsection{Liquidity for advice-gathering from and influence-delegation to specialists}
	(focus on pure delegation here, but also
	forward-ref "voter understanding" subsection later)

\input{liquid/risk}

\input{liquid/rel}

}

\input{liquid/conc}

%% file: liquid/intro.tex
\section{Introduction}


Democracy is in the midst of a credibility crisis.
Some of the most well-established Western democracies
have become increasingly polarized~\cite{prior13media,iyengar15fear}
to the point of
tribalism~\cite{hawkins18hidden,packer18report}
and authoritarianism\cite{browning18suffocation}.
The information sources voters use to understand the world
and make their decisions is increasingly suspect~\cite{woolley16automating,ferrara16rise,woolley17computational,broniatowski18weaponized,shao18spread}.
While democracy preaches a gospel of treating all citizens as equal,
established democracies fail to protect the equality
of citizens' influence at the ballot box~\cite{smith14political,gilens14theories,cost15republic,flavin15campaign,kalla16campaign,samuel18rigging,tisdall18american}.
\xxx{ voting rights discrimination, money-driven politics, gerrymandering}.

\com{
American politics today requires a word as primal as “tribe” to get at the
blind allegiances and huge passions of partisan affiliation. Tribes demand
loyalty, and in return they confer the security of belonging. They’re badges of
identity, not of thought. In a way, they make thinking unnecessary, because
they do it for you, and may punish you if you try to do it for yourself. To get
along without a tribe makes you a fool. To give an inch to the other tribe
makes you a sucker.~\cite{packer18report}
}

Outside the ballot booth, people in real democracies depend on government
to protect not only their physical safety,
but also their economic and social equality and human rights.
Here too, established democracies
fail to protect their citizens from private coercion
or feudal rent-seeking structures~\cite{shlapentokh11feudal}.
They fail to ensure equal access to equal economic opportunity
by accelerating transfers of public wealth
to the already-rich\xxx{hacker05abandoning,zupan17inside,stewart18republicans}
in the face of skyrocketing economic inequality~\cite{keller15partisan,piketty17capital},
fail to offer an adequate social safety net
to protect the ability of the unlucky or disadvantaged
to participate in society as equals with dignity, \xxx{safety-net}
and even fail event to protect many people from
effective slavery~\cite{weitzer15human,kara17modern}.
As Robert Dahl asked:
``In a political system where nearly every adult may vote but where knowledge,
wealth, social position, access to officials, and other resources are unequally
distributed, who actually governs?''~\cite{dahl61who}

\xxx{
To function, authoritariam depends only on human social behavior,
and a leader's ability to exploit it.	\xxx{cite}
Democracy, in contrast, must constantly struggle {\em against} human behavior,
depending on human institutions and many supporting technologies,
to protect citizens' rights, freedoms, and equality, and to function at scale.
\xxx{in the face of imperfect and self-interested voters and leaders alike}
\xxx{brief examples}
}

\subsection{Can Technology Revolutionize the Process of ``Rule by the People''?}

Today's democratic processes and institutions
were designed around assumptions rooted in paper-based bureaucracy,
that every interaction between people in which government is concerned
is costly both in human time
(people physically going to government offices and filling out forms)
and economically
(the costs of printing paper forms and hiring white-collar bureaucrats
to handle them correctly).
The main objective and optimization constraint in government
by in-person interaction and paper-based bureaucracy
is to minimize frequency of interactions
and to maximize what is accomplished by each.

Today's increasingly-pervasive networked computing technologies,
however, may hold the potential to reduce the cost of interactions
by many orders of magnitude:
enough to enable a qualitative ``phase change''
in applicable approaches to designing and building democratic institutions.
When interactions between people or with governments can happen anywhere,
at any time, with a button press or touch-screen gesture, 
requiring neither physical presence nor paper form-filling,
the feasible design space changes completely,
just as completely different processes and technological tools are applicable
when building a stone wall versus filling a swimming pool with water.

\subsection{Liquid Democracy: Essence, Origins, and Analogies}

This is the technology context in which {\em liquid democracy} arose:
stated vaguely and informally,
the idea that technology could free democracy from the clunky constraints
of paper ballots and government bureaucracies,
and enable voters to guide and direct their ``power of the people''
more easly, flexibly, and fluidly, like the flow of a liquid.
The term has no precise or standard definition,
and even its origin is unclear.
The specific term ``liquid democracy''
seems to have made its first recorded appearance on a long-defunct wiki
by a user going by the handle ``sayke''
and now preserved only on the
Internet Archive~\cite{sayke03liquid,sayke03voting}.
Most of the ideas associated with liquid democracy
were suggested earlier in various forms,
however~\cite{dodgson84principles,heinlein66moon,tullock67toward,miller69program,lanphier95model,ford02delegative}.

Since there is no single clear, standardized definition
of what liquid democracy actually means,
we will focus here on what the term {\em might reasonably} mean,
based on the namesake analogy of physical liquid.
As a physical state of matter,
liquid has two fundamental distinguishing properties:
it has no fixed shape but is {\em able to flow}
(like a any fluid including a gas),
and it is largely {\em incompressible} or volume-preserving
(unlike a gas). \xxx{cite}
Important properties derived from these fundamental characteristics
include that
liquids can be subdivided into nearly-arbitrary fractional portions
(treating their molecular limits as small enough
not to matter for most purposes),
and they may be stored and directed at low cost and effort
(via containers, channels, tubes, etc.).

As the purpose of any government
is to manage the flow and expression of {\em power}
(whether political, economic, or social),
the term {\em liquid democracy} naturally
suggests an approach to democratic governance
that manages expression and use of power like a ``liquid'':
\ie, a virtual substance whose flow people may direct or subdivide
easily at fine granularity and low cost.
The liquid democracy concept originally and most naturally applies to 
the nature of voting and democratic choice,
and that will be the focus of this paper.
The liquid analogy might also be applicable to
to other critical aspects of democratic governance,
such as ways voters obtain and vet information to make decisions,
and ways to protect equality in the social and economic opportunities
citizens need for effective participation.
We leave the exploration of these more far-reaching applications
of the liquid analogy to further exploration elsewhere, however.
\xxx{cite when appropriate}

In focusing on liquid democracy applied to democratic choice,
we will attempt to separate and analyze step-by-step
several of the entangled ideas of how liquid democracy
could make voting more ``liquid.''
We will attempt to disentangle and explore in some detail
the key ideas embodied in many of the variants of liquid democracy.
We will end up with something close to the idea of delegative democracy
that Ford proposed in 2002~\cite{ford02delegative},
but unpacking the ideas it contains step-by-step
and relating them to relevant precedents in existing democratic practices.
We will explore, in particular:

\begin{itemize}
\item
how the nearly-arbitrary {\em subdivisibility}
of a liquid applies to election systems
that allow voters to split and spread their voting power
among multiple alternative choices or candidates
(Section~\ref{sec:liquid:choice:spread});

\item
how a liquid's ability to flow may help us visualize --
and perhaps improve --
election systems that try to avoid ``wasted votes'' via vote transfer
(Section~\ref{sec:liquid:choice:trans});

\item
how the idea of guiding and directing a liquid's flow
suggests both old and new mechanisms to simplify voter choice
by {\em delegating} democratic voting power
to parties, organizations, individuals, or even algorithms
(Section~\ref{sec:liquid:choice:deleg});

\item
how both subdividing and guiding liquid voting power in combination
suggests solutions to the limits of
voter attention and enlightened understanding
that current limit the scalability of direct democratic participation
(Section~\ref{sec:liquid:spec});
and finally,

\item
how a liquid approach might make the {\em timing} of deliberation
and democratic choices more fluid
and give citizens more effective control over the democratic agenda
(Section~\ref{sec:liquid:choice:time}).
\end{itemize}

The purpose of this paper is not to analyze
any of these possible applications of the ``liquid'' analogy in great depth,
but rather to take a high-level perspective
on how they {\em might} be useful and potentially fit together.

Making innovative changes to decision structures,
or other elements of democratic governance processes,
inherently present risks.
We explore some of these risks in cases where they are readily apparent,
but make no pretense at having exhaustively identified all such risks.
More detailed formal or experimental analysis of the ideas of this paper
remains for future work.
Technology ultimately holds both great promise and great peril for democracy;
the purpose of this paper is to focus on the former
while acknowledging the latter.

%% file: liquid/spread.tex
\section{Liquidity in Enriching Choice by Spreading Vote Power}
\label{sec:liquid:choice:spread}

Even if the desired collective {\em outcome} of an election
is a single winner,
this does not necessarily imply that each voter's {\em input} to the election
must -- or should -- necessarily be just one single choice,
despite that being the most common practice.
A voter's preferences may be more nuanced.
For example, a voter may prefer first-choice candidate $A$,
but be willing to live with less-desired alternatives $B$ or $C$,
while truly despising candidate $D$.
If $A$ has little chance of winning,
then the voter must often choose between expressing her true preference for $A$
and ``wasting'' her vote,
or strategically trying to help more-electable candidates $B$ or $C$
win over $D$.
Similarly, if a majority would prefer either centrist candidate $B$ or $C$
but their support is split,
then an extremist candidate $A$ or $D$ might win
with a relatively small plurality of support
despite being least-desired by a majority of the electorate.

Recognizing these strategic conundrums,
many election systems have been devised
that allow voters to express support for multiple candidates,
in effect ``spreading'' their voting power
instead of lumping it into an all-or-nothing choice.
Since one of the basic properties of a physical liquid is
the ability to be subdivided arbitrarily,
viewing voting power or support as a liquid may be
a useful and interesting way to understand
voting systems that allow voters to split or spread their choice.
This section explores several existing vote-spreading schemes
in this light,
using the liquid analogy to illustrate their operation.

\xxx{ address score voting too at some point,
	as pure mechanism to measure strength of preference/opinion? }

\subsection{Approval Voting: Vote Spreading at No Cost}

{\em Approval voting} asks voters not to make a single choice
but instead to make a yes-or-no ``approval'' decision
on each candidate individually.
Voters effectively choose an arbitrary {\em subset} of the candidates
they consider ``above the bar'' or meeting whatever threshold they set,
without expressing any preference among those they approve.
The vote is tallied simply by adding the number of approvals
each candidate receives
and choosing the candidate with the highest approval score.
If the voter prefers $A$ but can live with $B$ or $C$,
for example,
then she can approve all three in order to help any of them win against $D$.

By allowing voters to support both a most-prefered candidate
together with at least one realistically-electable candidate,
approval voting allows voters to avoid ``wasting'' their vote.
Approval voting is also often seen as desirable
because it tends to prefer ``centrist'' candidates,
who may have weak support from a majority,
over ``extremist'' candidates who have strong support of a minority
but little support in the rest of the electorate.
Finally, approval voting is attractive because it allows paper ballots
to be laid out exactly as with traditional single-choice ballots --
\eg, with a checkbox or oval next to each candidate --
but merely stating in the instructions
that voters may choose multiple candidates.
\xxx{ cites }

\begin{figure*}[t]
\centering
\begin{tabular}{cc}
\subfloat[Example of an empty ballot in its initial state
	before the voter has allocated ``voting liquid''
	to any issues or candidates.
]{\includegraphics[width=0.30\textwidth]{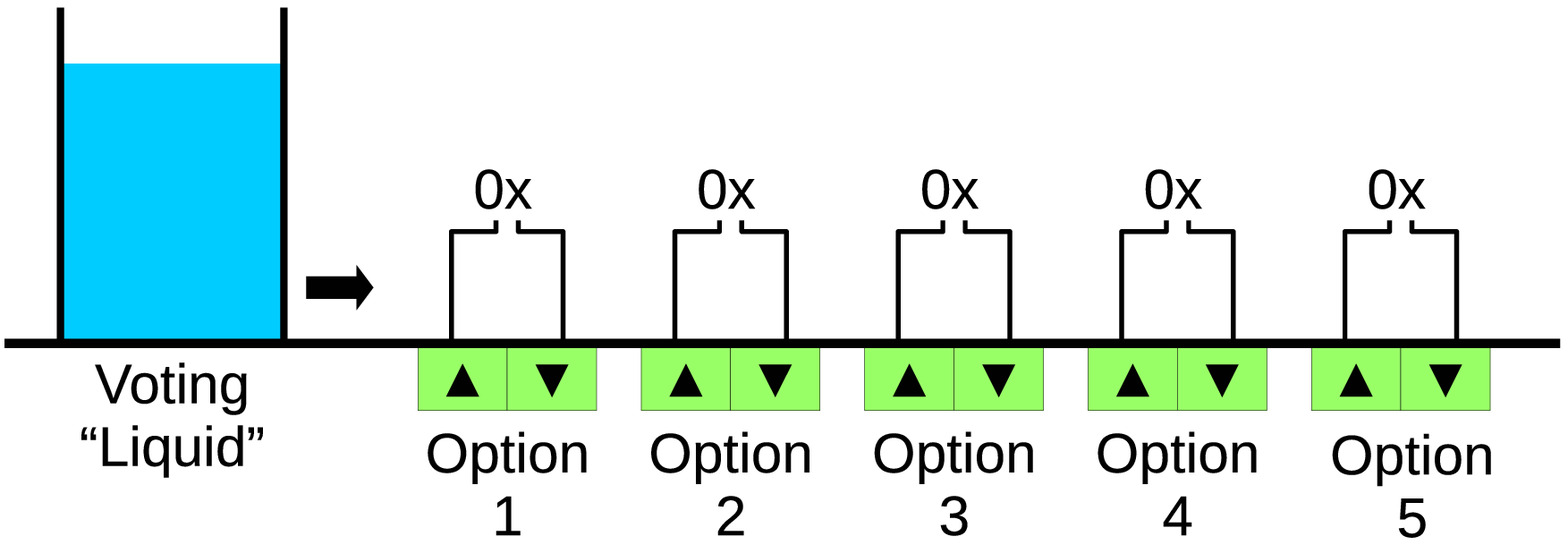}}
&
\subfloat[Example of a hypothetical voter who expresses no preference
	by supporting {\em all} the available options.
	The net effect is the same as an empty ballot.
]{\includegraphics[width=0.30\textwidth]{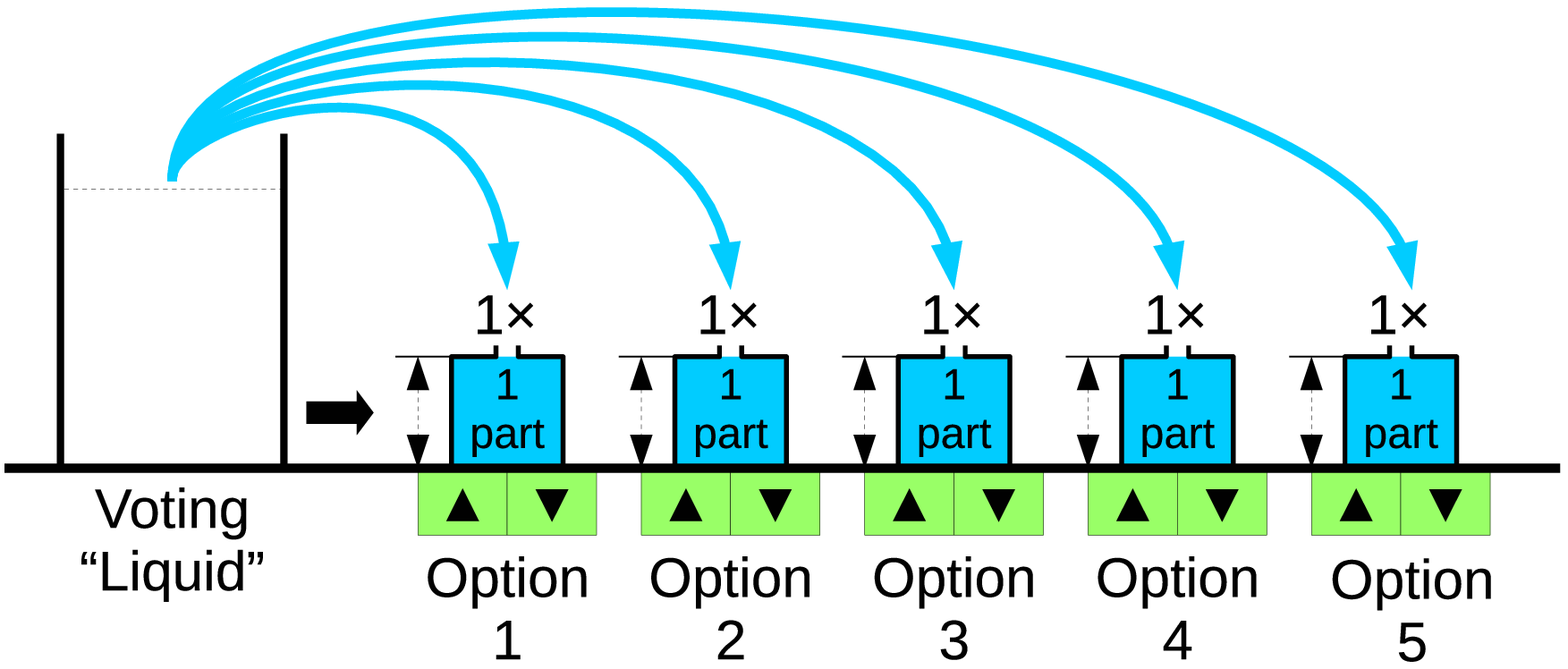}}
\end{tabular}

\begin{tabular}{ccc}
\subfloat[Example of a voter who clicks only one up-arrow
	to approve only one option.
]{\includegraphics[width=0.30\textwidth]{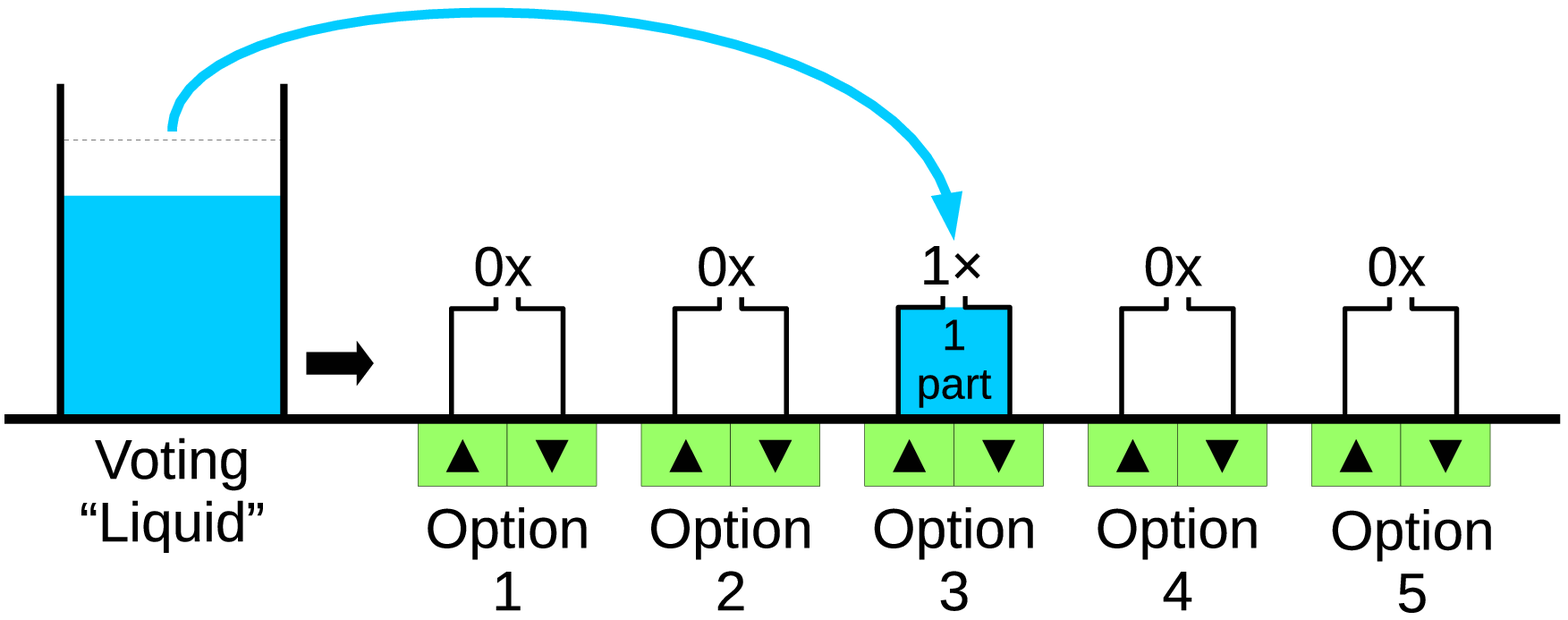}}
&
\subfloat[Example of a voter who approves two of five options.
]{\includegraphics[width=0.30\textwidth]{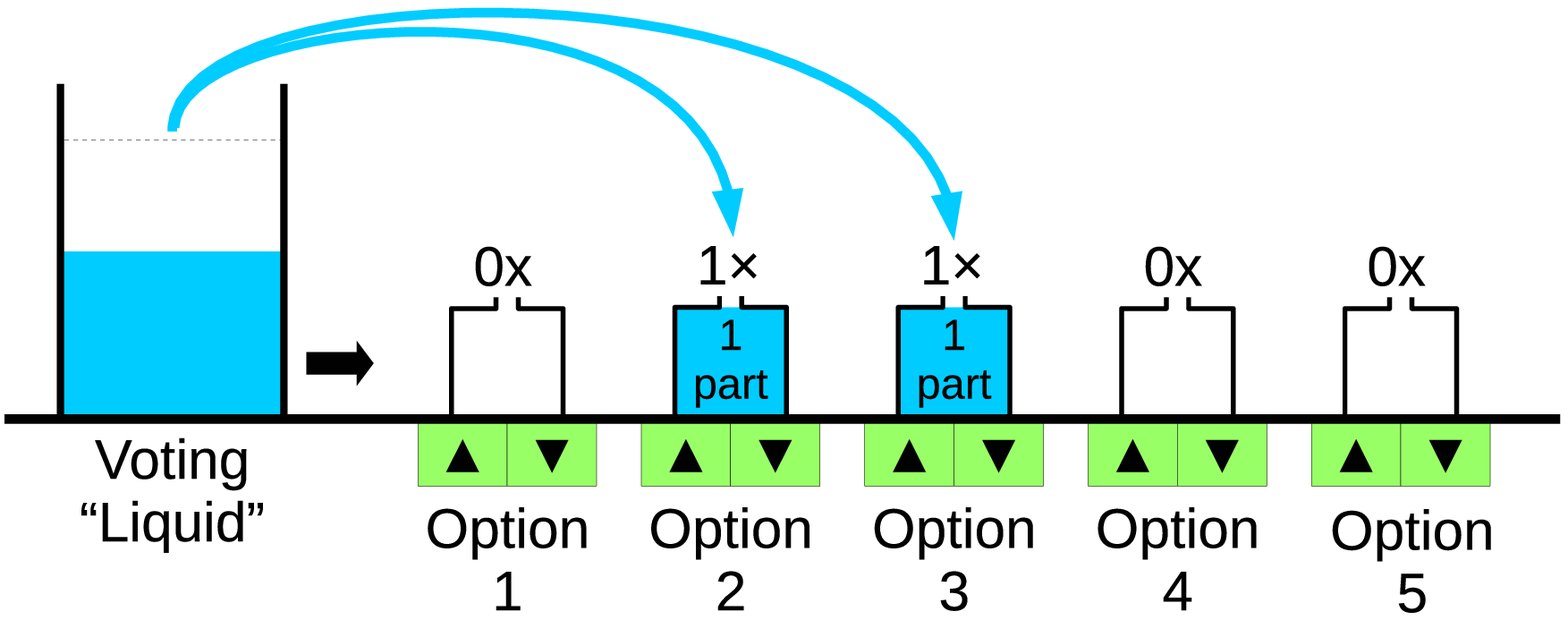}}
&
\subfloat[Example of a voter who approves all but one option.
]{\includegraphics[width=0.30\textwidth]{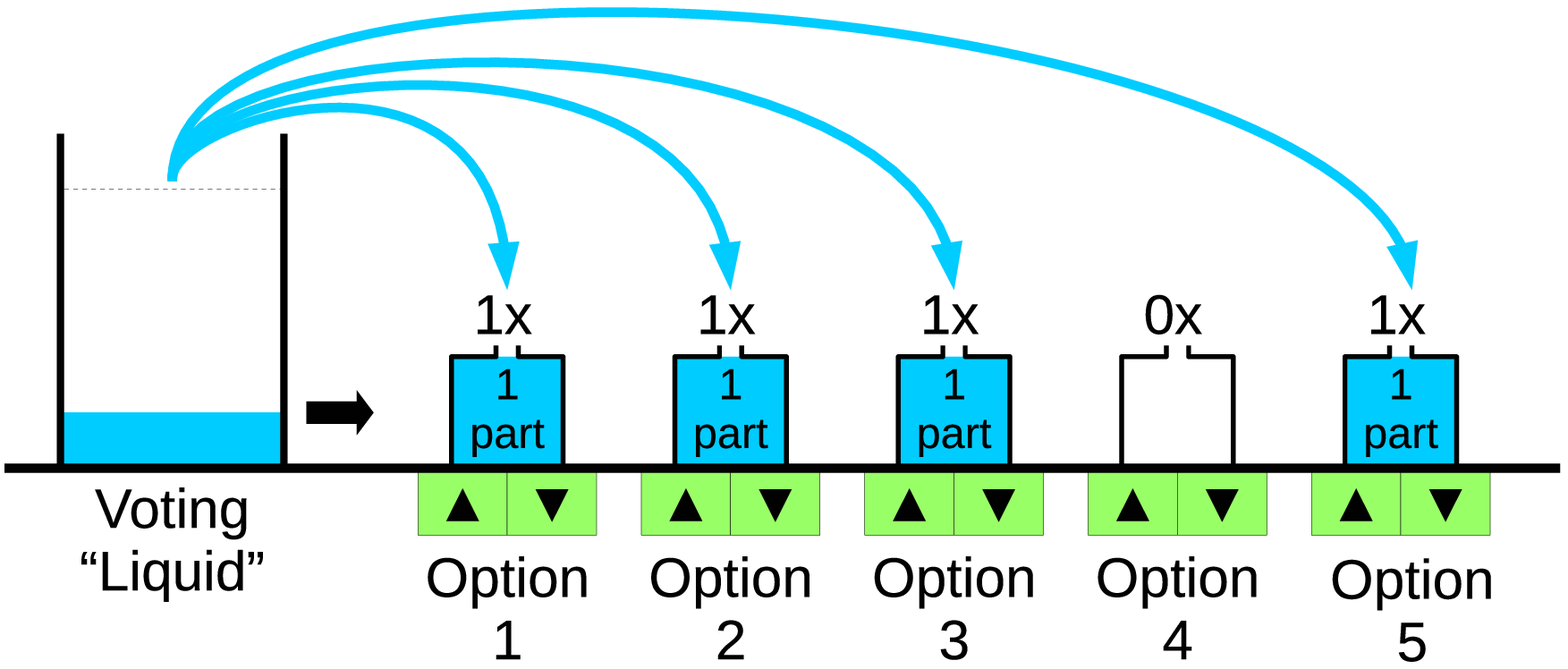}}
\end{tabular}

\begin{tabular}{c}
\subfloat[Example ``liquid tally'' of the three example ballots
	(c), (d), (e) above.
]{\includegraphics[width=0.45\textwidth]{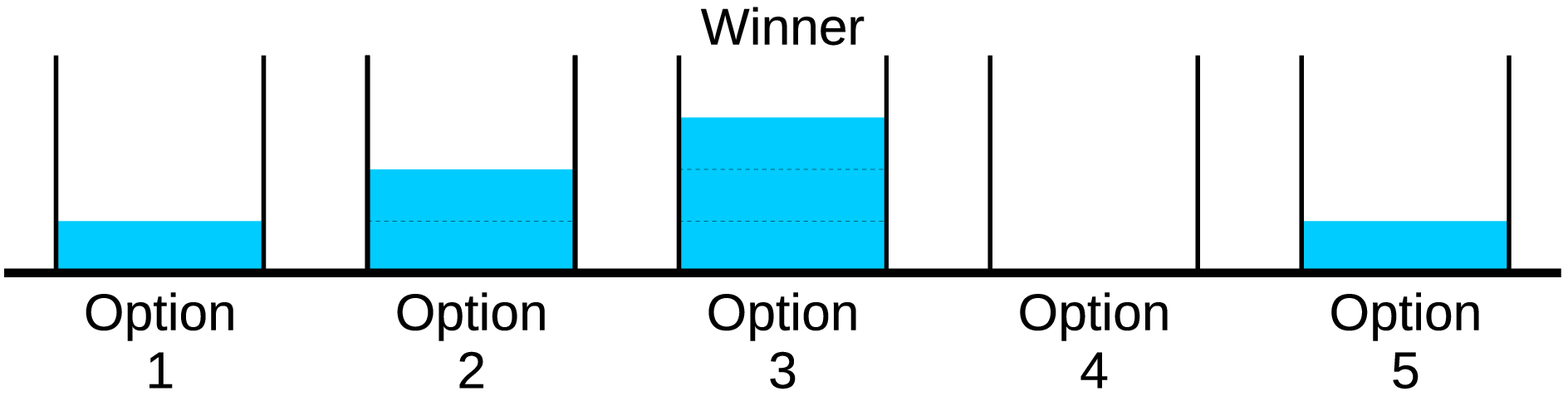}}
\end{tabular}
\caption{Liquid approval voting examples}
\label{fig:liquid:approv}
\end{figure*}

While approval voting is certainly simple enough
that we don't ``need'' a liquid analogy to understand or explain it,
we nevertheless take it as a starting point for exploring
the applicability of the liquid analogy to vote-spreading systems.
As illustrated in Figure~\ref{fig:liquid:approv},
we imagine each voter to have a pitcher of ``voting liquid,''
which the voter uses to fill (or leave empty)
each of a set of fixed-size containers, one for each candidate.
Each voter starts with enough liquid to fill all candidates' containers,
if desired --
although that particular choice (b) is equivalent to casting no vote
or filling no containers,
since it ``helps'' all candidates equally
and hence helps none relative to the others.
Thus, in practice each voter will have some unused liquid left-over
after approving a proper subset of the candidates.
The ``tallying'' process in this visualization
simply amounts to collecting all the liquid cast for each candidate
from the respective containers in all the voters' ballots,
and comparing each candidate's total amount of collected liquid (f).

By giving each voter ``enough'' voting liquid
to choose any subset of candidates
without affecting the amount of liquid conferred to each one,
approval voting effectively allows voters to spread their vote
with {\em no cost or penalty} for choosing more candidates:
\ie, it allows vote spreading with {\em no scarcity}.
In subsequent sections
we will use this liquid analogy to contrast this system
with related approaches that {\em do} impose a cost 
on choosing more candidates.

One critique of approval voting is that it requires voters
to divide candidates into just two ``bins'' (approved or unapproved),
while offering no obvious principle for how
that arbitrary approval ``bar'' should be set.
Further, voters have no way to express strength of preferences
among the subsets of candidates either above or below that bar.
Variations such as score voting \xxx{cite}
and majority judgment voting~\cite{balinski11majority}
address this issue by allowing voters to ``grade'' candidates on a scale
(\eg, A, B, C, D),
at a cost of increased complexity and less-familiar ballot structures.
Finally, although there are adaptations of approval voting
to multi-winner elections~\cite{kilgour10approval,brams14satisfaction},
they tend to be complex,
often difficult to tally even with computers~\cite{aziz15computational},
and involve seemingly-arbitrary vote ``re-weighting'' functions.

\xxx{
Note that approval voting has the appeal of using
the conventional ballot structure
while collecting {\em exponentially more} information
from each voter about each question.
Based on Shannon information theory~\cite{XXX},
a ballot question allowing voters to pick one out of $n$ choices
collects only $log_2(n)$ bits of information,
whereas a question allowing voters to pick {\em any subset} of $n$ choices
collects $n$ bits of information.
}

\subsection{Cumulative Voting: Vote Spreading with Economic Scarcity}
\label{sec:liquid:choice:spread:cum}

Consider now the long-established precedent of {\em cumulative voting},
a technique still commonly used in corporate governance processes
such as board elections. \xxx{cite}
Instead of assigning only one vote to each voter
(or to each voting share in a corporation),
the voting authority typically assigns some equal number $V$ of votes per voter
(or per share).
This approach enables voters not only to spread their voting power
among multiple candidates but also to express relative preferences between them.
Voters have a choice not only of {\em whom} to vote for,
but also of {\em how much} (\ie, what percentage)
of their total voting power to assign each candidate.
If a voter likes candidate A twice as much as B,
she can cast about two-thirds of her voting power to A and one-third to Bob.
Voters may still choose to assign all their voting power to one candidate,
a strategy known as {\em plumping}.	\xxx{check, cite}
On the other hand, if a voter has 10 votes to cast
and likes two candidates about equally,
she can cast five votes for each of those two candidates.

A cumulative vote in essence acts like a specialized ephemeral currency,
in which each voter receives an equal number of ``coins''
to ``invest'' in one or several candidates as they see fit.
Cumulative voting thus follows a conventional economic scarcity principle:
a coin (vote) spent on one candidate cannot also be spent on another.
Besides enabling voters to express strength of preference,
cumulative voting is often seen as a way to protect minority interests
in multi-winner elections in which the top $k$ candidates win seats,
because a minority coalition holding a $1/k$ fraction of total voting power
can use a plumping strategy to win at least one representative seat.
Cumulative voting does not solve the ``wasted votes'' issue, however:
a voter who casts even just one of several votes or coins
for a candidate with no realistic chance of being elected
reduces the voter's ultimate influence over
which of the more viable candidates get elected.


While cumulative voting allows voters to express strength of preference,
it usually requires them to do so at a fixed granularity
set by the voting authority.
Voters must ``calculate fractions'' to translate their preferences
into a suitable number of votes to give each candidate.
The globally-determined number of votes per voter $V$
may not be evenly divisible by the fractions reflecting the voter's preferences,
complicating the ``mental arithmetic'' demanded of each voter.
If a voter supports candidate A twice as much as B, for example,
and would thus like to cast 2/3 to A and 1/3 to B,
but the number of votes $V$ per voter is (say) 10,
then the voter must round: \eg, to 7 votes for A and 3 for B.
This rounding introduces error in the voter's expression of preferences
(about 5\% in this example).
We can reduce error by assigning more votes per voter, of course:
\eg, $V=100$ allows the example voter
to assign 67 votes to A and 33 to B,
reducing error to 0.5\%.
This finer granularity clearly comes at a cost
of increased complexity in the mental arithmetic required of voters,
and likely increase of voter mistakes:
it is much easier for people to see immediately
that several one-digit numbers sum to 10
than to verify that several two-digit numbers sum to 100.

\com{
In essence, cumulative voting increases voter choice
by enabling subdivision of voting power --
a capability that becomes more fine-grained and ``liquid''
as the number of votes per person ($V$) is increased.
Of course, given that votes almost invariably
express noisy, subjective human judgments,
it is debatable how much precision in preference expression
people actually need,
and the benefit of increasing $V$
undoubtably reaches a point of diminishing returns quickly,
giving way to the costs of forcing voters
to think about or calculate multi-digit fractions.
}

\paragraph{Liquid Cumulative Voting:}

Here we encounter a situation in which viewing voting power as a liquid
may lead to interesting improvements on existing voting practices.
We tend to think of a physical liquid
as if it were {\em arbitrarily} subdivisible in any desired fractions:
although there is still a minimum granularity (\eg, one atom or molecule),
it is small enough not to matter for practical purposes.
People intuitively divide liquids by arbitrary fractions or ratios
in everyday practices as ancient as cooking:
mix two parts water with one part vinegar,
pour half into the pan.
Could the rich expression of voting preferences
be made as simple and intuitive as handling liquids in cooking?
While voting using real liquid would no doubt get messy,
could an electronic voting system with a suitable user interface
allow users to visualize their voting power as a virtual liquid,
and divide and spread it among candidaters
in whatever ratios make sense to them?

\begin{figure*}[t]
\centering
\begin{tabular}{cc}
\subfloat[Example of an empty ballot in its initial state
	before the voter has allocated ``voting liquid''
	to any issues or candidates.
]{\includegraphics[width=0.30\textwidth]{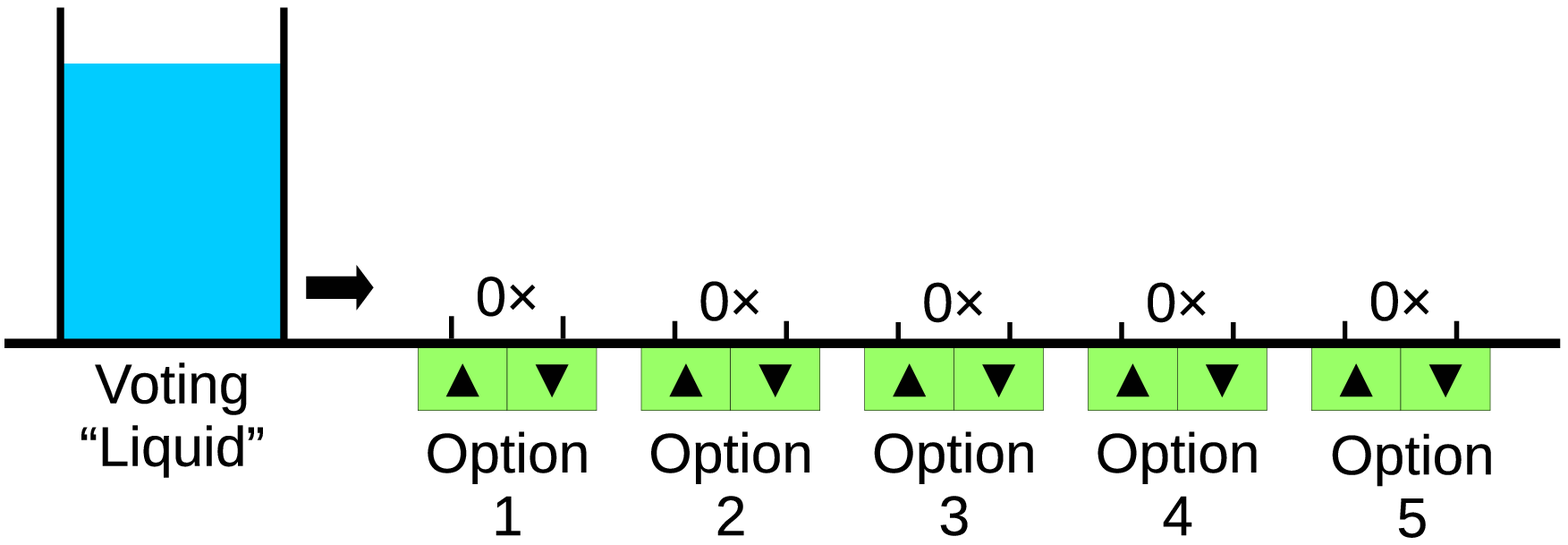}}
&
\subfloat[Example of a hypothetical voter who expresses no preference
	by allocating equal voting liquid to each of the options.
	The net effect is the same as an empty ballot.
]{\includegraphics[width=0.30\textwidth]{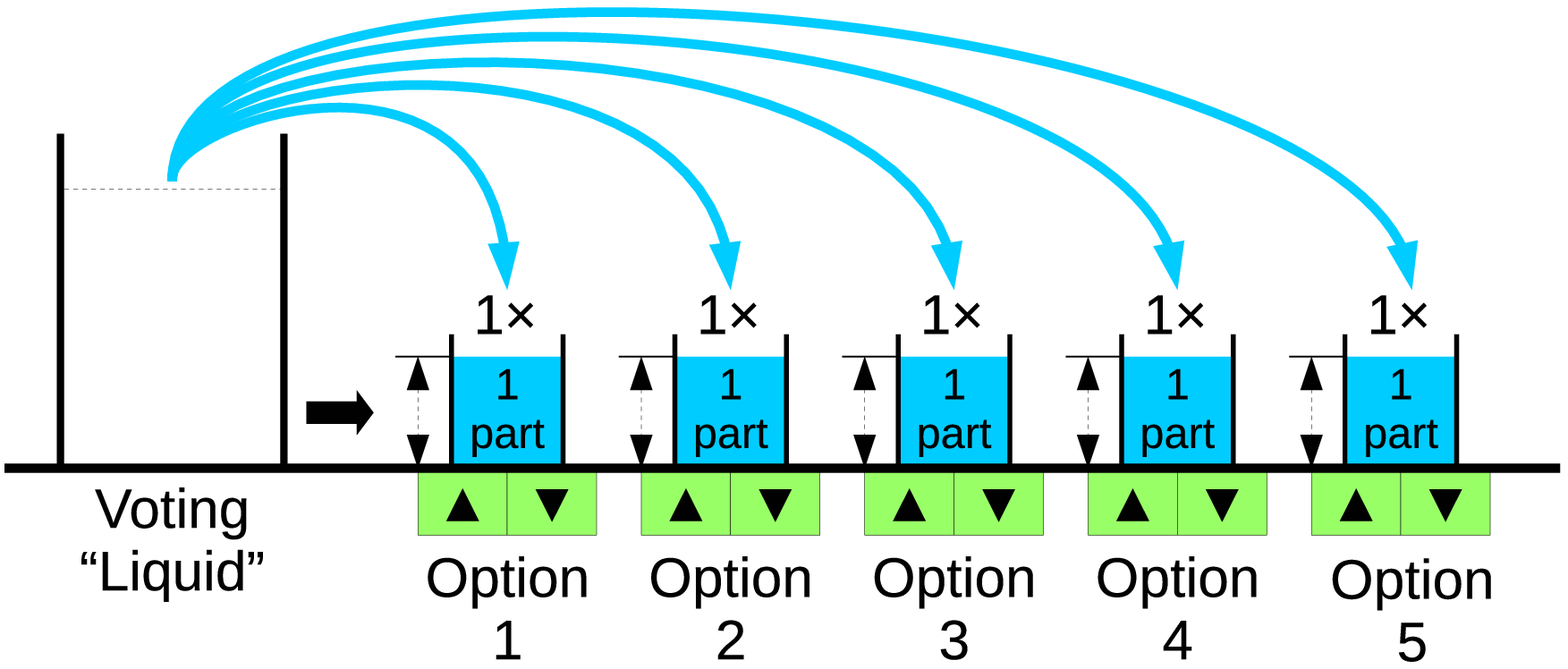}}
\end{tabular}

\begin{tabular}{ccc}
\subfloat[Example of a ``plumping'' voter who clicks only one up-arrow
	to place all her voting power on one option,
	using all the voting liquid in just one part.
]{\includegraphics[width=0.30\textwidth]{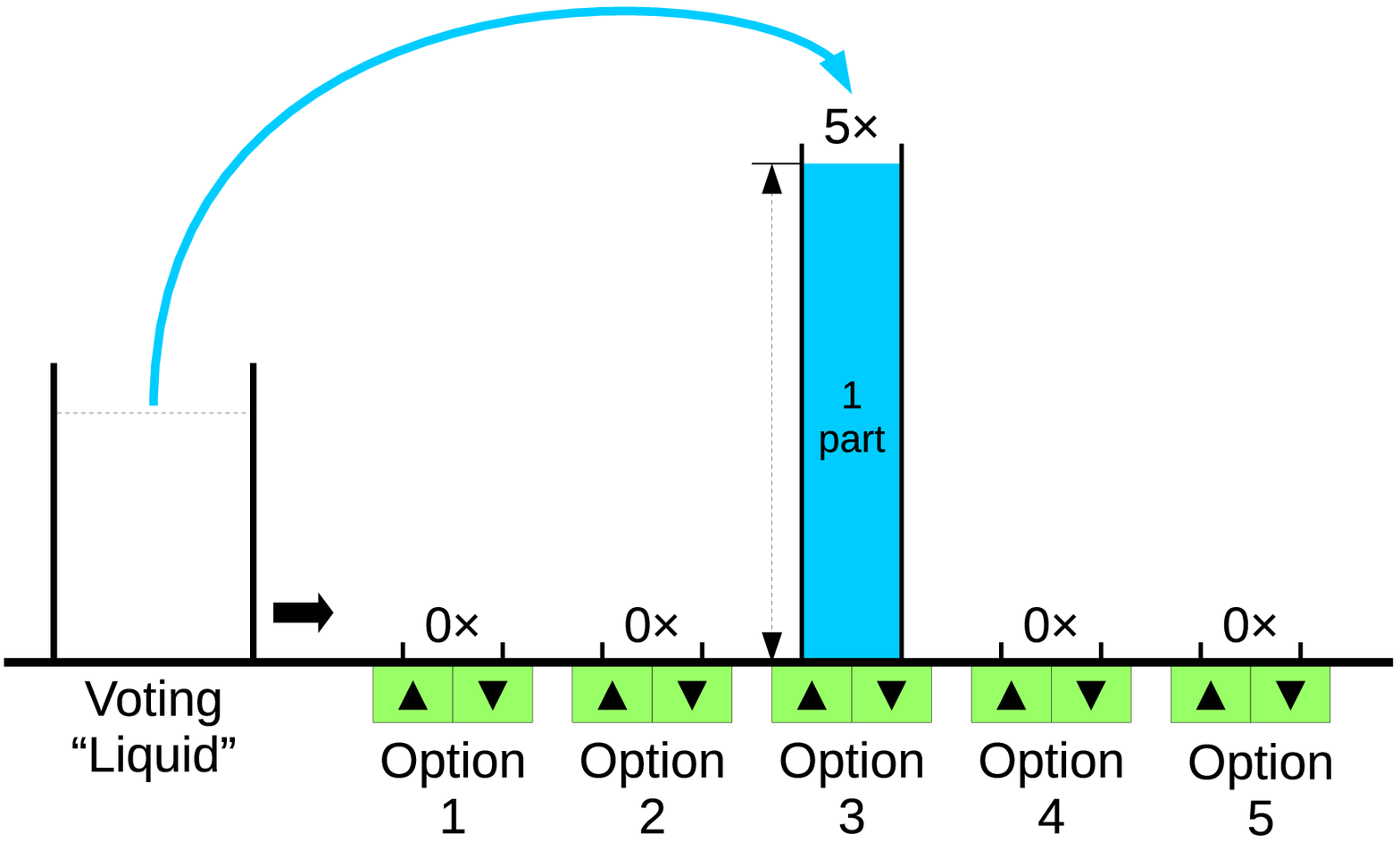}}
&
\subfloat[Example of a voter who supports two of five options,
	one twice as strongly as the other,
	dividing the liquid into three parts.
]{\includegraphics[width=0.30\textwidth]{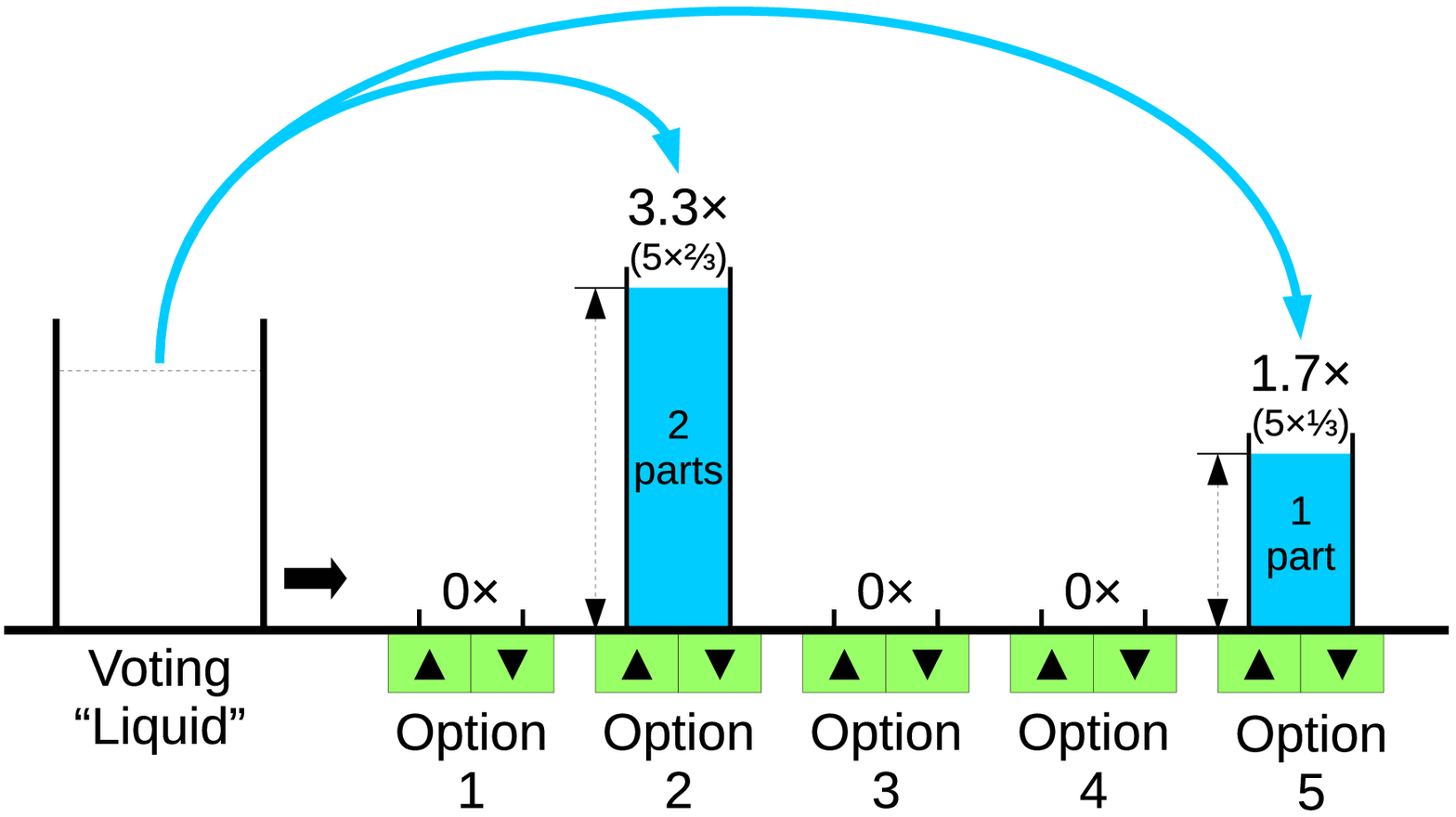}}
&
\subfloat[Example of a voter who supports {\em all but} one option,
	allocating the voting liquid in equal parts
	among the other four options.
]{\includegraphics[width=0.30\textwidth]{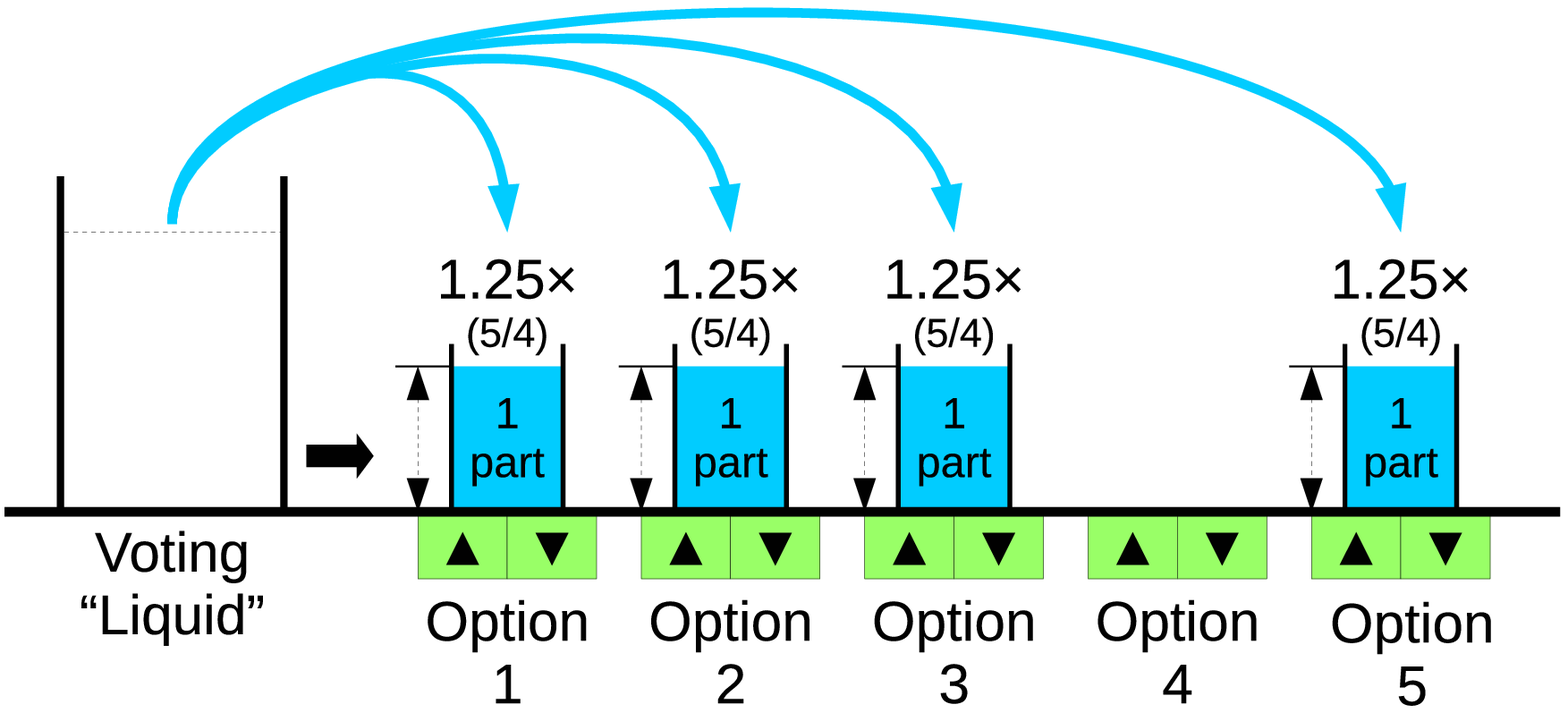}}
\end{tabular}

\begin{tabular}{c}
\subfloat[Example ``liquid tally'' of the three example ballots
	(c), (d), (e) above.
]{\includegraphics[width=0.45\textwidth]{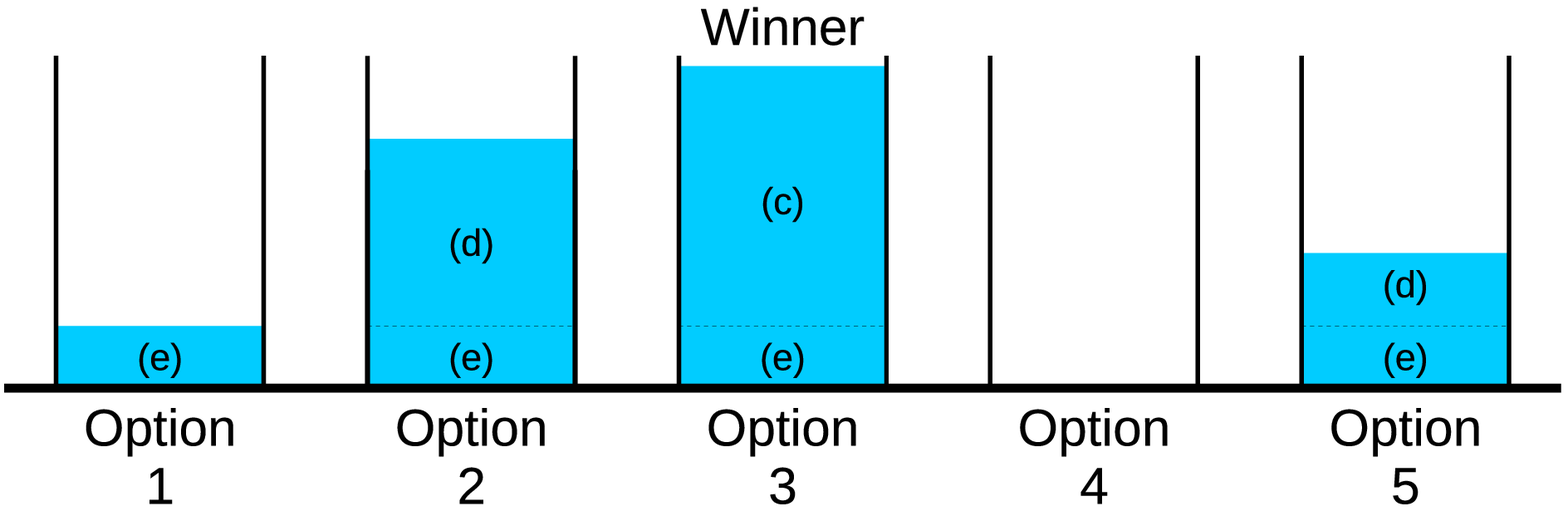}}
\end{tabular}
\caption{Liquid cumulative voting examples
	with ``subdivision by parts'' interface}
\label{fig:liquid:cum}
\end{figure*}

Figure~\ref{fig:liquid:cum} illustrates one way this might be accomplished.
Each user is given an equal amount of virtual voting liquid,
which they can divide into any number of equal ``parts''
for allocation to the candidates.
Clicking the up-arrow for a candidate adds one part to that candidate,
decreasing the proportion of liquid represented by each part.
In Figure~\ref{fig:liquid:cum}(b), for example,
the voter has created just one part,
assigning all his virtual voting liquid to option 3.
In Figure~\ref{fig:liquid:cum}(c),
the voter has assigned two parts to option 2 and one part to option 5,
expressing support for these two options,
the former twice as much as the latter.
Tallying the vote simply amounts to adding up all the liquid portions
from all voters according to candidate,
as illustrated in Figure~\ref{fig:liquid:cum}(f).

Voters no longer need to translate their preferences into coarse-grained units
arbitrarily determined by the voting authority (\eg, 10 or 100 votes each).
Instead, each voter independently decides
the number of parts to to divide his voting liquid into,
while fairness is preserved by the fact that each voter
gets the same amount of liquid.
In practice the electronic voting system may need to reduce
these voter-defined fractions at some point down
into some minimum-granularity unit for tallying,
but this granularity can be arbitrarily small
to ensure high precision (\eg, ``microvotes''),
without users ever needing to be aware of it.
\com{
It then becomes feasible, and arguably advantageous,
to increase the number of atomic votes $V$ assigned to each voter
to a large number, \eg, one million ``microvotes,''
to ensure that vote tallying is performed with high precision
regardless of how individual voters split their vote.
In the above example, the voter's assignment might
``behind the scenes'' automatically become
666,6667 microvotes for Alice and 333,333 for Bob.
The voter need never be aware of these atomic units of microvotes,
except in the rare case she wants to dig into the e-voting system's
technical details.
}

\xxx{Bitcoin analogy: Satoshis are extremely small,
but no one needs to think in or calculate with Satoshis.}

\begin{figure}[t]
\centering
\includegraphics[width=0.40\textwidth]{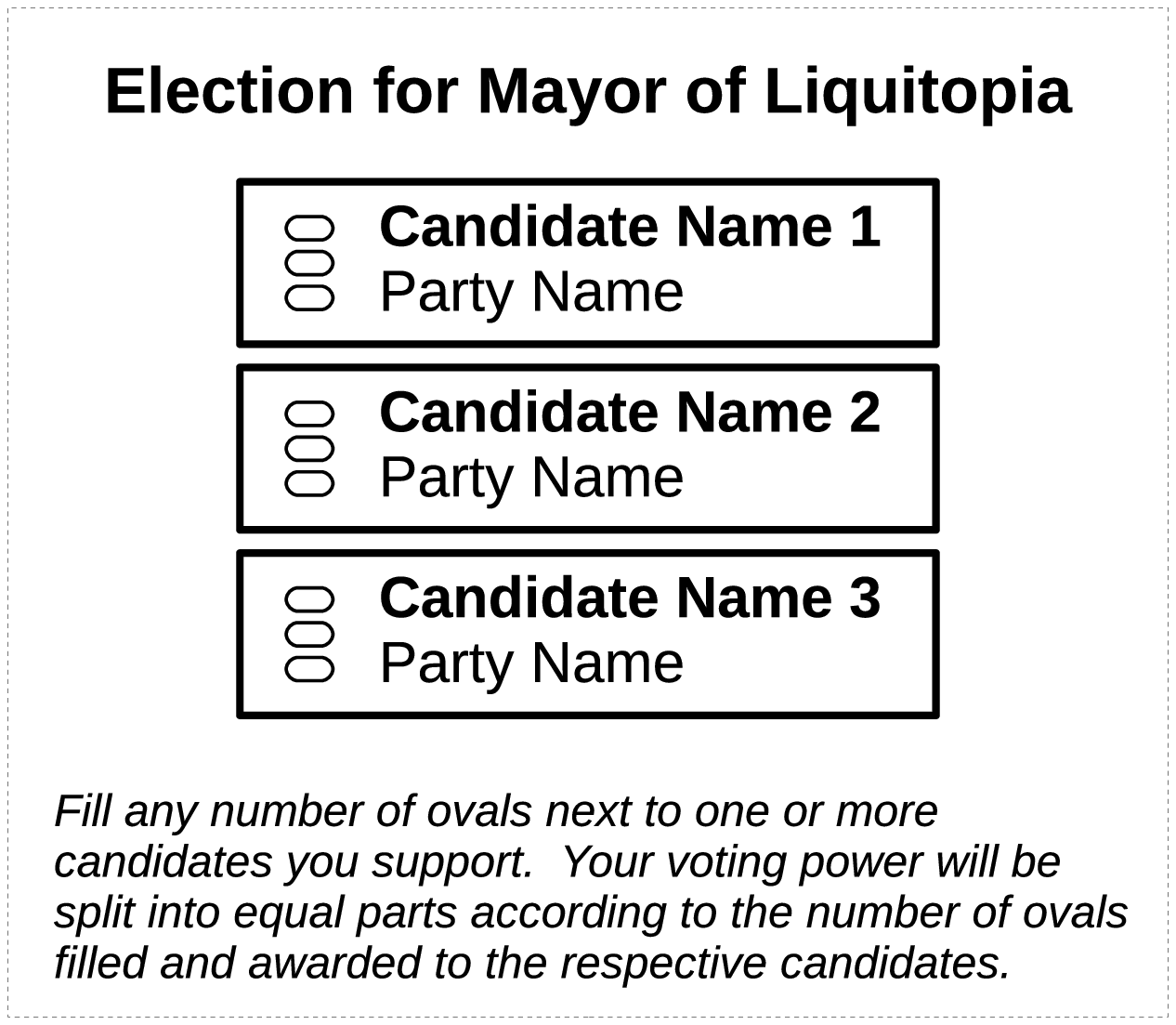}
\caption{Example paper ballot for liquid cumulative voting}
\label{fig:liquid:cum-paper}
\end{figure}

While an electronic ``liquid simulation'' voting interface
like in Figure~\ref{fig:liquid:cum}
might be a powerful and intuitive way to allow users
to subdivide their vote by parts,
a similar effect could be approximated with conventional paper ballots.
Consider the example ballot in Figure~\ref{fig:liquid:cum-paper},
which looks exactly like a conventional paper ballot
except with several fillable ovals next to each candidate instead of just one.
The ballot instructs voters to fill {\em any number of ovals}
next to whichever candidate or candidates they support.
The voter's total voting power is divided into equal parts
according to the total number of ovals filled,
and distributed to the candidates in those proportions.
With this approach, it remains trivial for voters to ``plump''
all their voting power onto one candidate as in conventional voting:
simply fill in one oval (or all)
next to the single most-preferred candidate.
But voters can also spread their voting power by intuitive ratios,
\eg, filling two ovals for A and one for B,
to express that they support both but consider A twice as desirable.

\subsection{Quadratic Voting: Vote Spreading with Attenuated Cost}
\label{sec:liquid:choice:spread:quad}

\xxx{	Connect my quadratic voting discussion to Julia Cage's article
	and her discussion of it. }

The recently-proposed idea of quadratic voting~\cite{posner15voting,lalley18quadratic}
ties voting to economic theory even more closely.
Like cumulative voting,
quadratic voting assumes each voter has a pool of voting ``coins'' or tokens:
perhaps apportioned in equal measure to each voter,
in proportion to the amount of stock held in a company,
or even purchased directly for real money,
depending on the variation.
Also like cumulative voting,
voters can express strength of preference
by spending all their voting tokens on one candidate or issue,
or spread it among multiple different candidates or issues.

In quadratic voting, however,
the voter pays the {\em square} of the number of votes cast
for or against a given candidate or issue:
\eg, casting one vote costs one coin,
but casting two votes for the same candidate or issue costs four coins.
Thus, casting more votes for a given candidate or issue
costs not just more coins but progressively {\em more coins per vote}.
In a certain rational model of voting behavior,
there is an argument that quadratic voting incentivizes voters
to reveal their true strength of voting preferences,
by being willing to pay more per vote on candidates or issues
they care more about,
and little or nothing on choices for which they have weak or no preference.

\begin{figure*}[t]
\centering
\begin{tabular}{cc}
\subfloat[Example of an empty ballot in its initial state
	before the voter has allocated ``voting liquid''
	to any issues or candidates.
]{\includegraphics[width=0.30\textwidth]{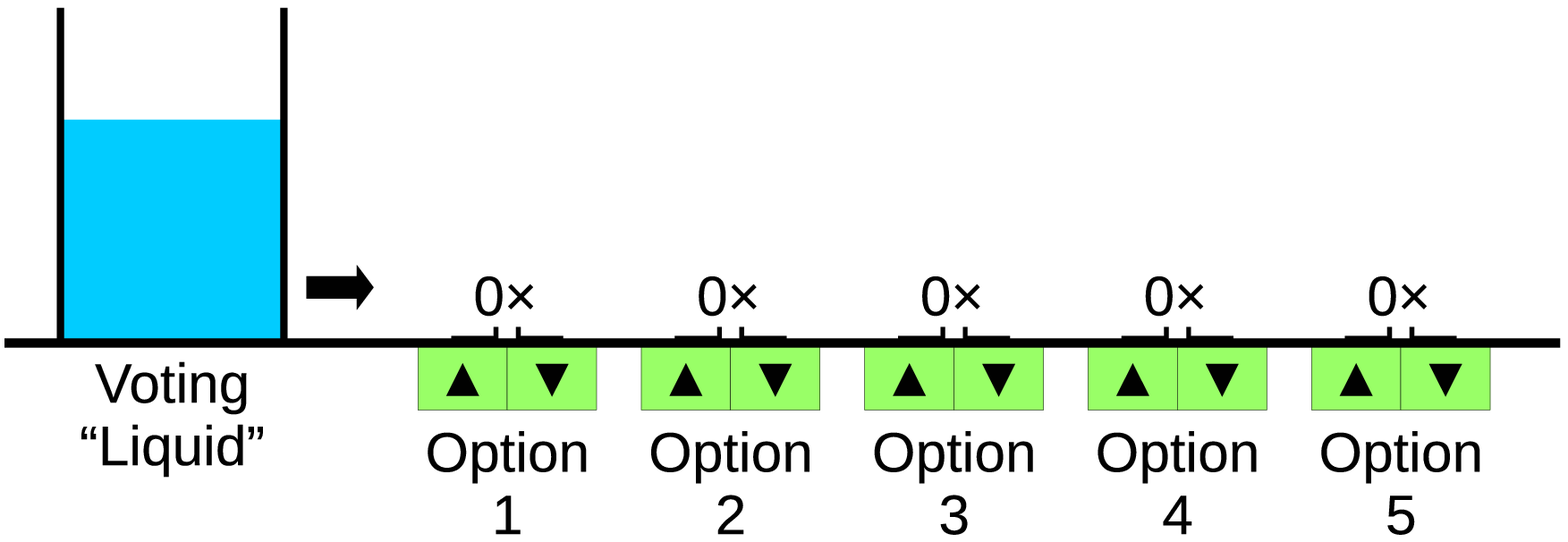}}
&
\subfloat[Example of a hypothetical voter who expresses no preference
	by allocating equal voting liquid to each of the options.
	The net effect is the same as an initial/empty ballot.
]{\includegraphics[width=0.30\textwidth]{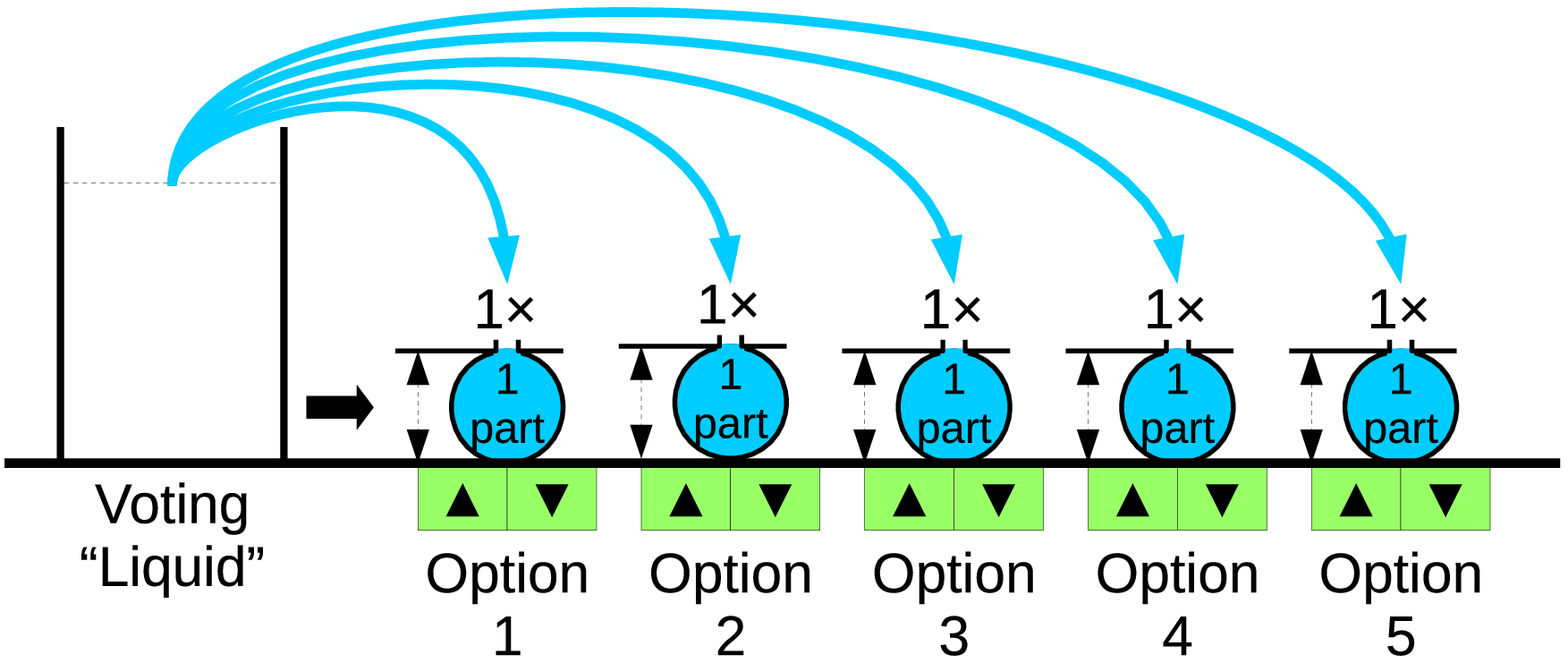}}
\end{tabular}

\begin{tabular}{ccc}
\subfloat[Example of a voter who supports only one option,
	using all five parts of voting liquid to cast $\sqrt{5}$ votes.
]{\includegraphics[width=0.30\textwidth]{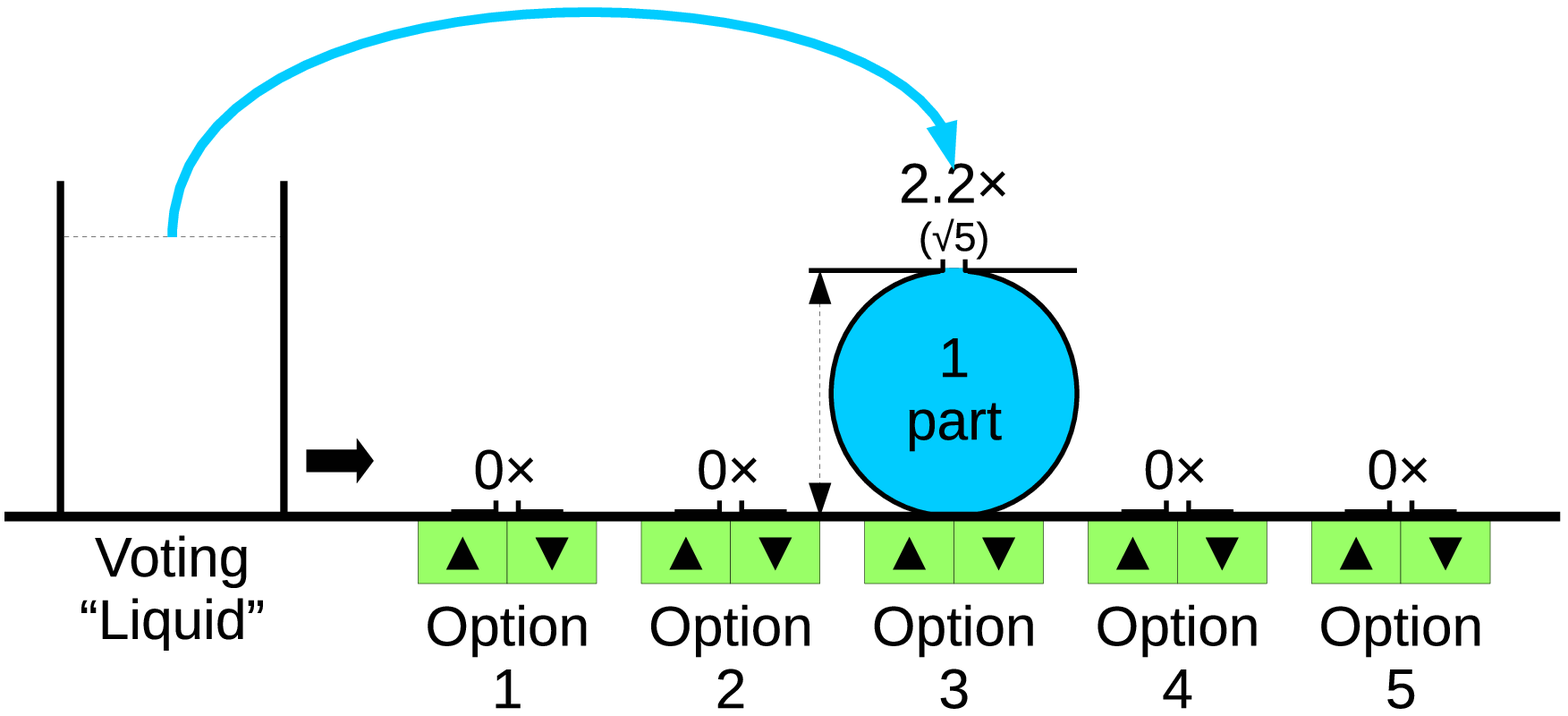}}
&
\subfloat[Example of a voter who supports two options, one strongly
	using four parts of liquid to cast two votes,
	the other weakly
	using one part of liquid to cast one vote.
]{\includegraphics[width=0.30\textwidth]{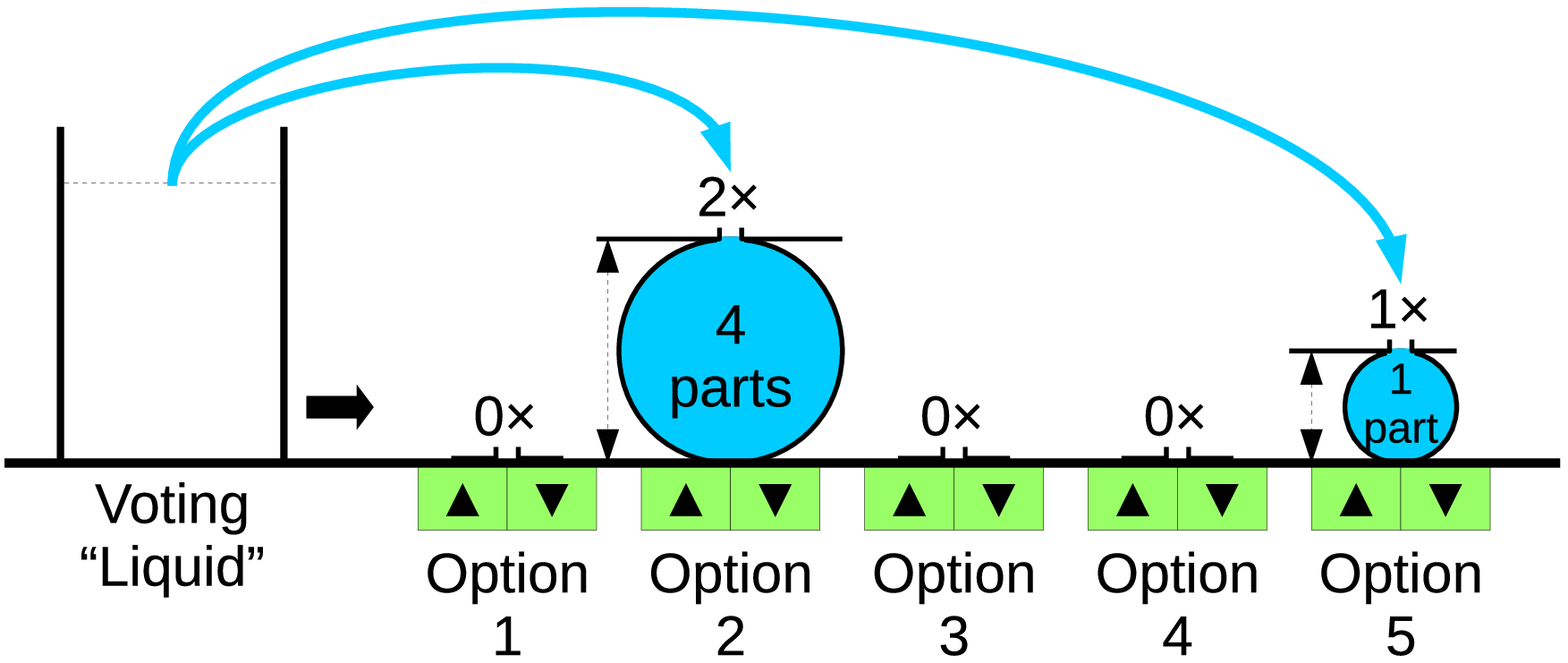}}
&
\subfloat[Example of a voter who supports {\em all but} one option,
	allocating the voting liquid in equal parts
	among the other four options.
]{\includegraphics[width=0.30\textwidth]{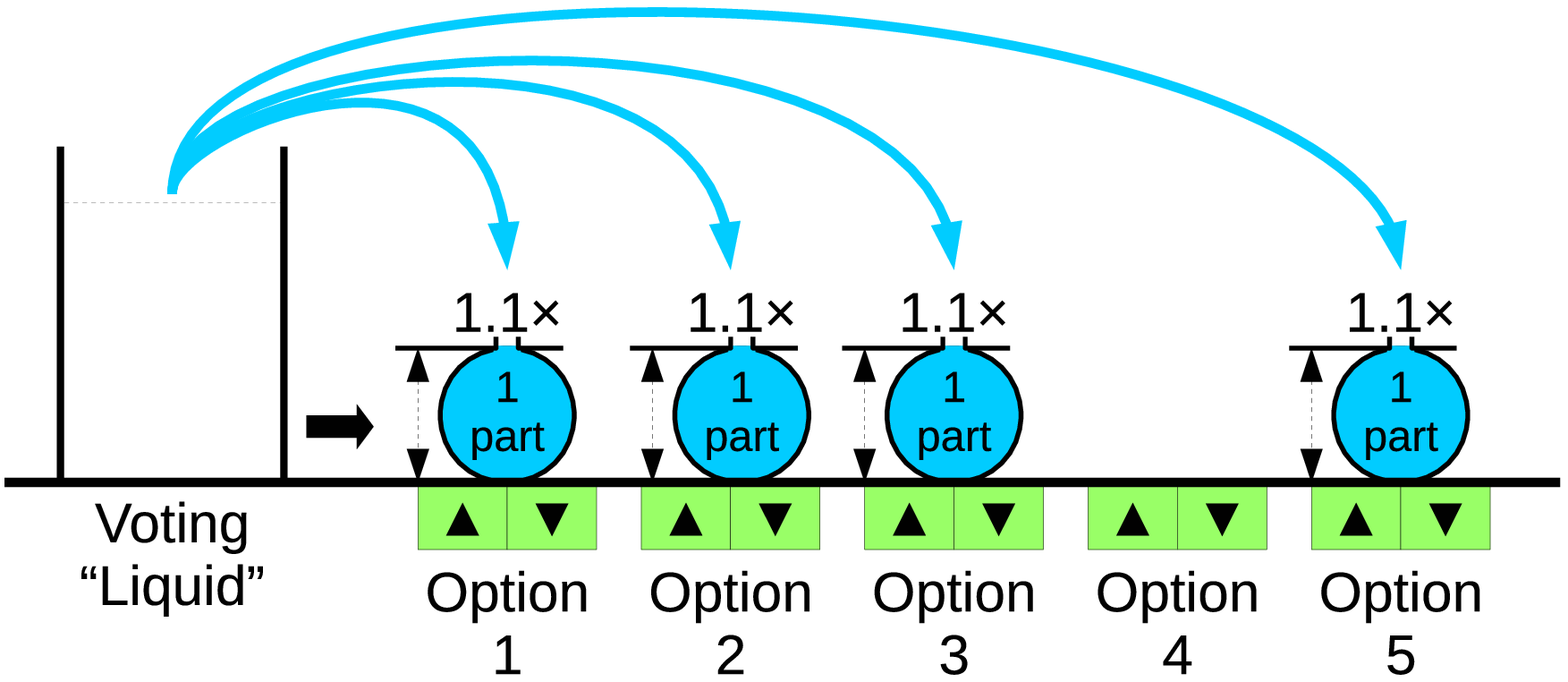}}
\end{tabular}

\begin{tabular}{c}
\subfloat[Example tally of the three example ballots
	(c), (d), (e) above.
]{\includegraphics[width=0.45\textwidth]{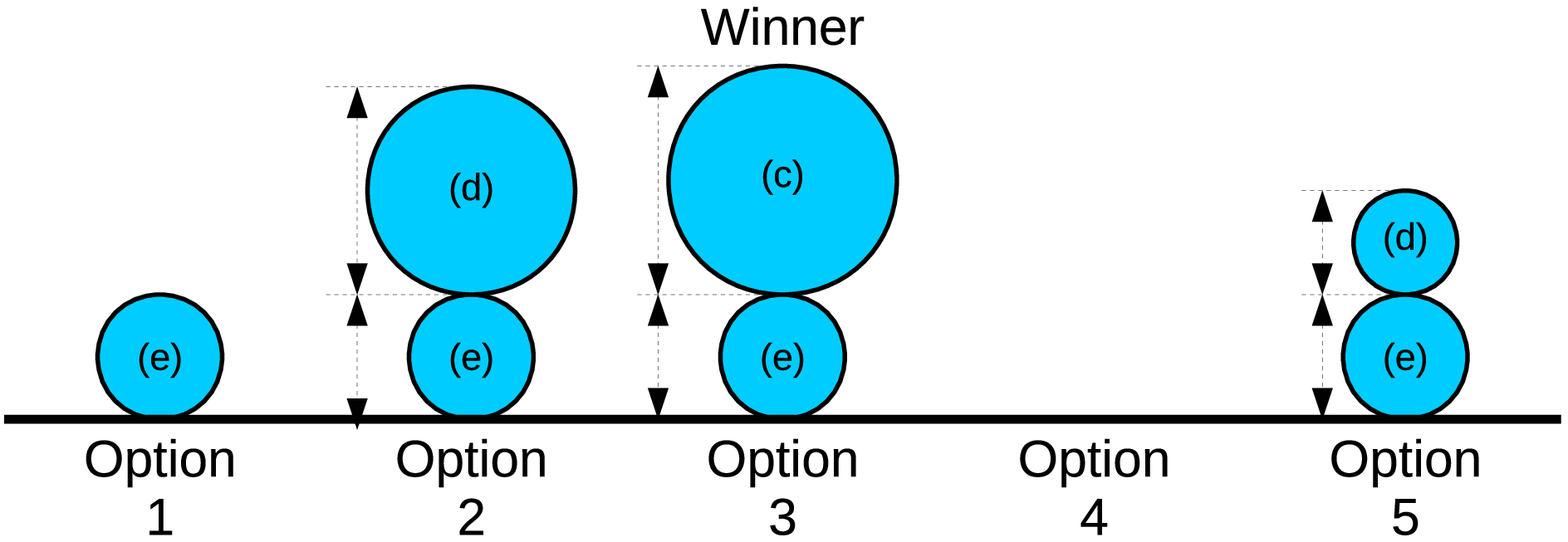}}
\end{tabular}
\caption{Liquid quadratic voting examples}
\label{fig:liquid:quad}
\end{figure*}

In practice,
many ordinary human voters are likely to be unaware
what ``quadratic'' even means,
let alone understand
the sophisticated incentive-compatibility argument behind it --
or make the complex probabilistic strategy calculations
of the ideal rational voter it assumes.
The liquid voting power analogy might offer a way
to make quadratic voting slightly more intuitively comprehensible,
however.

\paragraph{Liquid Quadratic Voting:}

Consider Figure~\ref{fig:liquid:quad},
illustrating a {\em liquid quadratic voting} analog
of the liquid cumulative voting example above.
As before, each voter is given an equal amount of virtual voting liquid,
which they may divide into any number of parts
and assign to candidates or issues.
However, in this case the ``containers'' they are filling
are not vertical beakers as above,
but rather elastic ``water balloons'' in two-dimensional space.
Adding more liquid to one candidate's balloon
expands its two-dimensional {\em area} proportionately,
but it is the balloon's {\em height} (circular diameter)
that determines how much the vote affects the corresponding option.
As illustrated in Figure~\ref{fig:liquid:quad}(f),
the resulting votes are tallied by ``stacking'' the water balloons
(in a perfect virtual world with no squashing due to gravity)
and measuring the total height of each candidate's stack.

Viewed in this way,
quadratic voting represents an interesting intermediate point
between approval and cumulative voting.
Recall that approval voting imposes {\em no cost} to the voter
for approving more candidates rather than fewer:
voting liquid is not ``scarce'' at all,
and each approved candidate is ``helped'' the same amount
regardless of how many other candidates the voter also approves.
Cumulative voting imposes a high cost on vote spreading:
\eg, a voter who expresses equal support for two candidaters
will help each of those candidates half as much as a plumping voter.

Quadrative voting, in contrast,
effectively incentivizes vote spreading by imposing 
{\em some} cost -- but a moderate, attenuated cost --
on supporting more options.
A voter who plumps all of his liquid on only one candidate
maximizes the amount he helps {\em that single candidate}
but loses ``cost-efficiency'' in doing so.
A voter who spreads his liquid among two or more candidates
helps each candidate less but maximizes the {\em aggregate} amount
he helps {\em all} the candidates he supports.
In Figure~\ref{fig:liquid:quad}, for example,
voter (c) uses all of his voting liquid to help option 3,
maximizing the amount he helps that option
($2.2\times$ the baseline balloon height in (b)),
whereas voter (e) provides less help to each of the four option he supports
but while helping all of them more in aggregate ($4.4\times$ the baseline).

\xxx{ missing detail: quadratic voting as proposed
allows voters to ``buy'' votes to vote either for or against an option,
so in case (e) the voter could actually assign all his liquid
to a negative vote for option 4 with height $\sqrt{5}$. }

\paragraph{Risks of Coercion and Illegal Vote-Buying:}

While QV has certain appeal, it also comes with risks.
For example,
QV assumes that voters exercise their free choice independently
and cannot collude with other voters outside the electoral system.
If voters can secretly coerce or collude with each other,
they can secretly buy or trade votes at only linear cost per vote
rather than the quadatic cost that QV intends to impose.
Suppose for example that Eve has four QV coins or tokens
with which to cast a vote for an issue or candidate she cares about.
If she votes honestly, she will be able to buy and cast two votes.
If she can secretly find four apathetic voters
willing to cast one vote each on her behalf
in exchange for an equivalent of one coin each on some black market,
then she can effectively cast {\em four} votes through these co-conspirators,
and sell the use of {\em her} four coins on the same black market,
exactly recovering her cost while doubling its effective power.

In practice it seems unlikely that any implementation of QV
could prevent secret coercion and collusion risks entirely,
short of pervasively surveiling all voters' interactions with each other
and eliminating their privacy and free choice in the process.
These risks could be mitigated by coercion-resistance mechanisms
such as ``receipt-freeness'' in properly-designed voting systems, however,
ensuring that if Eve tries to buy votes from co-conspirators,
she has no way of verifying that they ``stayed bought''
and actually cast the votes she paid them to.
For this reason, implementing proper coercion-resistance mechanisms
in systems implementing QV may be even more important
than in more traditional elections.

\xxx{ quadratic voting risks:
encourages voters *not* to participate or express an opinion (give-a-shit tax);
doesn't solve the attention scalability problem;
increases incentive for and hence risk of ``vote-buying'' attacks;
doesn't distinguish between voters who don't care because of ignorance
(it's important and effects them but they don't know it)
from those who have "enlightened understanding" but little/no preference;
risks making diversionary effects worse
(\eg, draw attention to hot-button social issues like guns and abortion
and drawing peoples' voting power away from economic topics like tax policy
that affect them more but they're less aware of)
}

\paragraph{Risks of Rewarding Apathy and Distraction Politics:}
\label{sec:liquid:choice:spread:quad:distract}

Another important risk stems
from the difference in information and understanding
between the ``perfectly-informed rational voter'' that QV theory assumes
and the decidedly-imperfect, often poorly-informed and non-rational character
of real-world voters.
In particular, QV's theoretical analysis makes no account
for the gap between the perfect rational voter's understanding
and a real voter's understanding of their situation:
\eg, the difference
between issues the voter doesn't much care about
because those issues in fact don't greatly affect him,
and issues the voter doesn't much care about
because he {\em doesn't understand} how much they affect him
when in fact they {\em do}.
A voter in the latter situation,
who lacks sufficient understanding of and appreciation for
the extent to which a given issue actually affects them in reality,
will ``underspend'' on that issue when making choices in a pure QV system,
investing little or nothing on this issue
and saving their coin for other issues she {\em knows} she cares about.

Thus, QV presents a risk of incentivizing and even rewarding voters
to be apathetic and not cast any vote on issues they know little about,
even if in reality it is in their best interests
to know about and vote on those issues.
It is hard enough in practice to get voters to show up and vote at all,
let alone ask them to {\em pay} --
even with special-purpose voting ``coin'' or liquid --
to vote on candidates or issues of vague and uncertain significance to them!

Further, deploying QV naively in public elections
could greatly compound the already-serious problem of
distraction politics~\cite{jamieson93dirty,weiskel05sidekick,leibovich15politics}.
To whatever extent powerful or moneyed interests
can distract voters' attention toward ``bright shiny objects''
such as political scandals,
divisive social issues (\eg, immigration)
or divisive moral controversies (\eg, abortion),
and cause voters to invest most of their voting power
in votes for or against these controversial candidates or issues,
special interests can thereby draw the bulk of the public's voting power
away from matters the special interests {\em actually} care about
and ensure that votes they care about meet little public resistance.
Whereas conventional distraction politics merely relies on voters
not paying close enough attention to notice and vote against special interests,
quadratic voting could give the practitioner of distraction politics
an even stronger {\em economic weapon}
with which to sap the voting power of potential public opposition.

\xxx{ cite Milner from deleg.tex on how two-party systems
	incentivize leaders to keep citizens less well-informed? }

%% file: liquid/trans.tex
\section{Liquidity in Proportional Representation via Transferable Voting}
\label{sec:liquid:choice:trans}

In addition to being subdivisible,
another important property of a physical liquid is that it {\em flows},
changing its position and shape easily
while preserving a nearly-constant total volume.
A liquid's flow can easily be directed (\eg, by channels or tubes),
much more easily than solids for example.
In viewing democratic voting power as a virtual liquid,
we can consider the ``flow'' property as
enabling users to direct their voting power in more flexible ways
while preserving preserving the democratic principle of equality,
meaning in this case that everyone has the same voting power
(the same volume of liquid)
regardless of what they do with it.
We can find precedent for legitimate transfer of voting power
in many well-established systems for democracy,
and in some cases visualizing these transfers of power as ``flows''
of a virtual liquid may potentially help people understand these systems.

One precedent for treating democratic voting power as liquid flows
may be found in transferable voting systems such as
Instant Runoff Voting (IRV) and Single Transferable Vote (STV),
discussed next.
IRV and STV ask voters not just to pick a single choice
but to rank their choices in preference order,
and automatically transfers votes down this sequence
as candidates are eliminated or elected in the vote-tallying process.

\subsection{Single-Winner Elections: Instant Runoff Voting (IRV)}

\xxx{	briefly summarize history somewhere:
	where STV and IRV originated.
Dodgson aka Lewis Carrol proposed concept of ``clubbing''
or delegation to avoid wasting votes in 1884.~\cite{dodgson84principles}
}

A basic, long-recognized problem with single-winner elections
in which there are more than two candidates or choices
is that none of the candidates might obtain a majority of the votes.
In fact it's quite possible for the candidate
{\em least favored} by a majority of the population to win,
if that candidate's minority support is focused
but the majority's vote is split among two or more similar candidates.
This is the source of the ``spoiler effect'' \xxx{cite Nader election etc},
the threat of which pressures voters to vote strategically
for a ``lesser of two evils'' instead of expressing their true preferences,
when their preferred candidate has little chance of winning.

This conflict between strategy and true expression of preferences
is the reason many countries schedule a separate {\em runoff} election
if no majority winner emerges from the initial multi-candidate election.
However, runoff elections are costly in many ways including
additional time required, election administration expenses,
and attrition of voter attention.

This problem motivated {\em instant runoff voting}, \xxx{cite}
which essentially asks voters to cast ballots only once
but provide on those ballots all the information the voting authority needs
to perform a virtual runoff election ``instantly,'' if needed,
without voters' further immediate involvement in each round.
The key is to ask voters to indicate not just first-choice candidates
but to {\em rank} any or all of the candidates in preference order.
After collecting these rank-order ballots,
the election authority uses them in a multi-round elimination process
in which the weakest candidate in each round is eliminated
and that candidate's votes transferred to next-choice candidates
on the respective voters' ballots.

\begin{figure*}[t]
\centering
\hfill
\subfloat[First round: voting power from all ballots
	flows to each voter's first-choice candidate.
]{\includegraphics[width=0.45\textwidth]{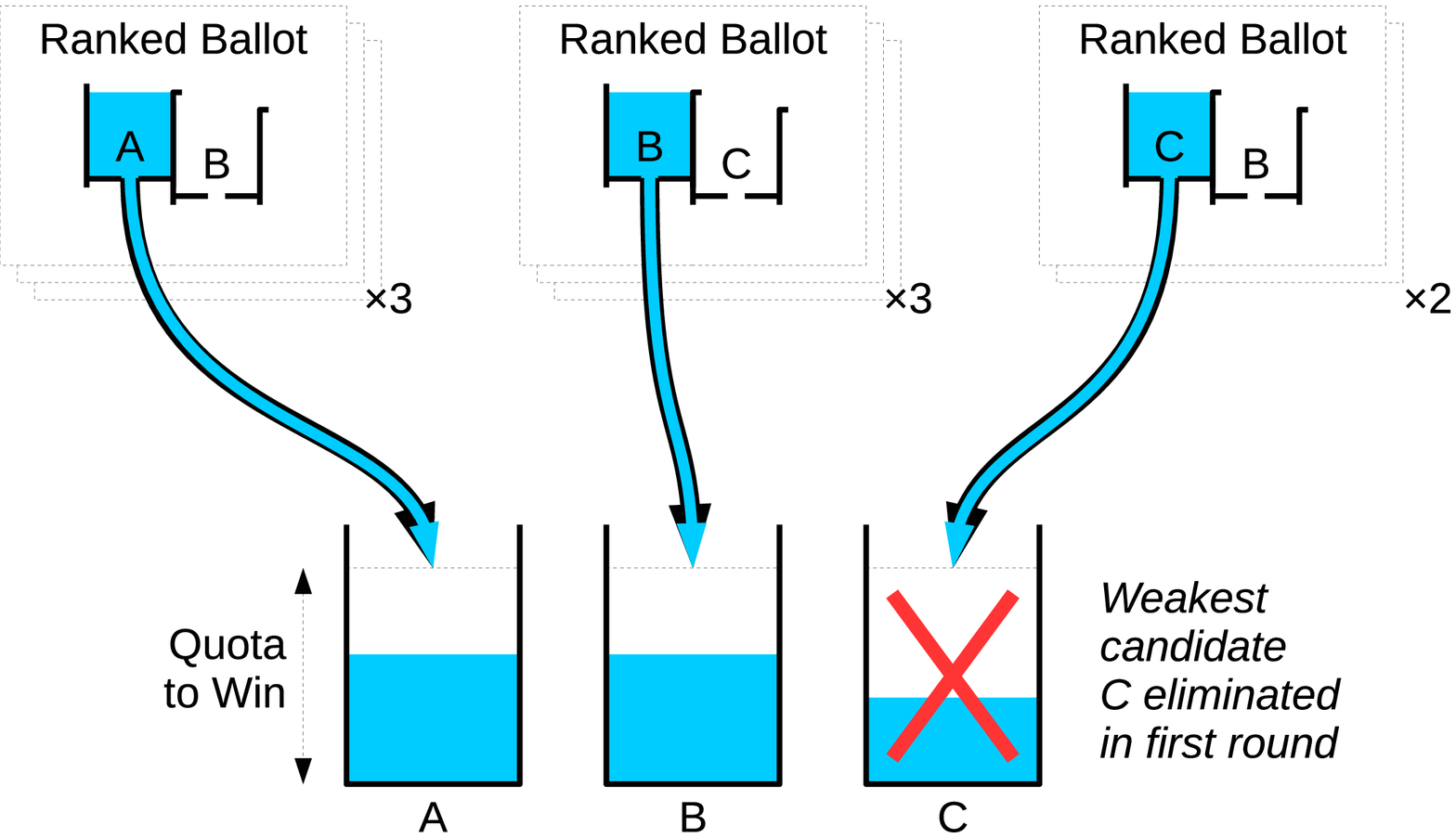}}
\hfill
\subfloat[Second round: weakest candidate $C$ eliminated;
	$C$'s voting power ``pumped'' back to voters' ballots,
	then ``flows'' (\ie, is transferred)
	to $C$ voters' second-choice candidates.
]{\includegraphics[width=0.45\textwidth]{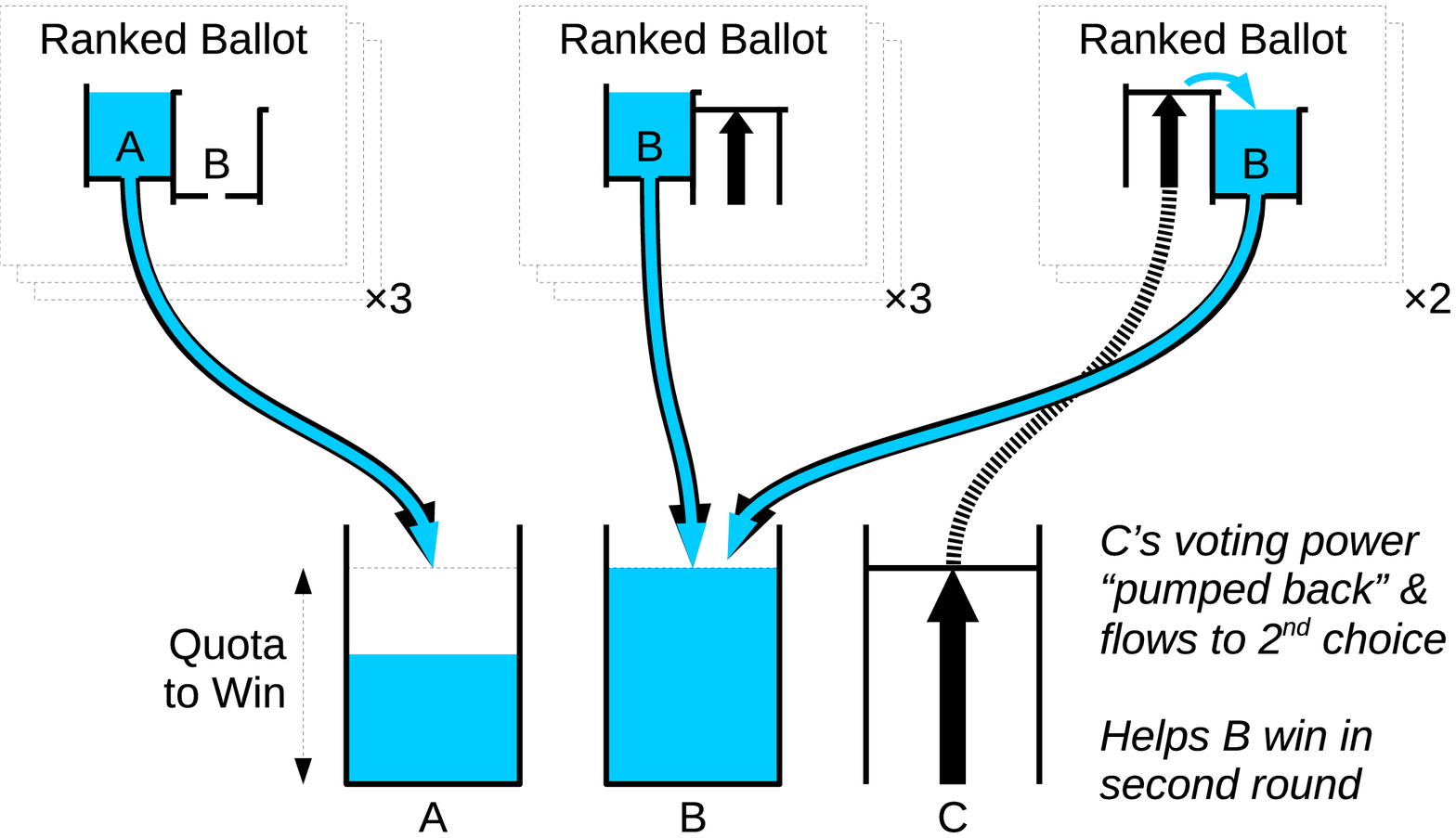}}
\hfill
\caption{Instant Runoff Voting (IRV) example illustrated
	via ``liquid flows'' of voting power}
\label{fig:liquid:irv}
\end{figure*}

One way to visualize this process of automatically transferring voting power
to next-choice candidates on a ranked ballot
is to consider voting power as a liquid whose flow can be guided and pumped,
as illustrated in Figure~\ref{fig:liquid:irv}.
In this simple IRV example,
eight voters total cast ballots with three different preference orders
($3\times$AB, $3\times$BC, and $2\times$CB).
Each voter's ballot starts with an equal allotment of imaginary voting liquid,
initially in a container representing the voter's first-choice candidate.
In the first IRV round,
each ballot's liquid flows to a larger container
that accumulates the aggregate ``political support''
of the voters' first-choice candidates.
If one of the candidates' ``fluid level'' reaches the quota line
representing majority support, that candidate wins the election.

In the Figure~\ref{fig:liquid:irv} example, however, there is no majority,
so the weakest candidate $C$ is eliminated.
Since $C$ no longer has a chance to win the election or make use of
the voting power conferred by $C$'s immediate supporters,
IRV ``pumps'' $C$'s accumulated voting liquid back to the supporters' ballots,
where the respective beakers for $C$ are likewise closed,
allowing each affected voter's liquid to flow to the next-choice candidate.
In this example, the two supporters of $C$ picked $B$ as their second choice,
enabling $B$ to reach a majority and win the election in the second round.

While the liquid analogy is by no means necessary or the only easy way
to explain and understand IRV,
it is appealing in several ways.
First, it directly embodies the basic democratic principles of
``rule by the people''
(liquid power flowing from voters' ballots to their chosen candidates)
and equality
(every voter having an equal volume of voting liquid,
which is preserved wherever and however it flows).
More specifically to IRV,
this analogy illustrates how a single-winner candidate can obtain
a majority of the population's delegated political power
(represented by $B$'s accumulation of a majority of total voting liquid),
and thus can lay claim to legitimate representation of the majority,
even if the population's first-choice votes are split among many candidates.
Finally, this visualization uses the capacity of liquid to ``flow''
in two important respects:
first in terms of explicitly representing the delegation of power
from ``the people'' to their representatives,
and second in terms of representing the automatic transfer of power
to next-choice candidates across rounds as weak candidates are eliminated.

The liquid analogy becomes even more relevant, however,
as we extend it to multi-winner elections.

\subsection{Multi-Winner Elections: Single Transferable Vote (STV)}

\xxx{ note somewhere that there is evidence proportional representation
succeeds in creating at least the perception that ``every vote counts''
by apparently yielding significantly higher
voter turnout~\cite{blais90proportional}. }

Single Transferable Vote (STV) is a generalization of IRV's principles
to multi-winner elections
in which there are both multiple candidates
and multiple seats to be filled.\xxx{cite STV}
STV's goals are not only to avoid votes being wasted,
but also to achieve {\em proportional representation},
ensuring that any sufficiently-large minority group
can elect a number of representatives
in approximate proportion to the group's size.

Like IRV, STV's goal is to encourage voters to express their true preferences
instead of being pressured to vote strategically,
and to minimize the number of ``wasted'' votes that do not
help elect (or contribute to the perceived legitimacy of)
a winning candidate in the end.
When there are not just one but $n$ seats to be filled,
STV observes that under suitable conditions it is possible to assure that
fewer than $1/(n+1)$ votes are ``wasted,''
a fraction that decreases toward zero (no waste)
as the number of seats $n$ increases.
To achieve this goal, however,
STV must address not just one but two reasons why votes might be ``wasted'':
first, as in IRV, votes for first-choice candidates not popular enough to win;
and second, unlike IRV, ``extra'' votes for choice candidates
who received {\em more votes than necessary} to win.

Consider a na\"{i}ve multi-winner extension to IRV, for example,
in which we simply eliminate the weakest candidates in succession,
transfering their votes to next-choice candidates according to the ballots,
and halt this process when there are $n$ candidates not yet eliminated.
Suppose that the most-popular first-choice candidate received
twice the number of votes needed to win a seat,
typically defined by the Droop quota: $Q = \lfloor{v/(n+1)}\rfloor+1$,
where $v$ is the total number of valid ballots. \xxx{cite}
In this case, half of this popular candidate's votes are effectively ``wasted''
in that they were not needed (in retrospect) to elect this candidate,
and they did not help elect
any other (perhaps closely-aligned) candidates either in the end,
because IRV only ever transfers votes from eliminated candidates.
It might have been strategically preferable for this popular candidate
to encourage some (up to half) of her voters to vote for {\em another}
allied but less popular candidate --
but this reintroduces the strategy conundrum,
and may be risky if the popular candidate has overestimated her support.

Mature STV systems, therefore,
transfer not only the votes of eliminated candidates
but also the ``extra'' votes of elected candidates,
so that a maximum number of votes eventually
apply toward electing {\em some} candidate to one of the $n$ seats.
In particular, STV follows an iterative process as in IRV,
but at each step we first check whether any current ``hopeful'' candidate
has passed the vote quota $Q$ needed to win a seat.
If so, we mark that candidate ``elected'' rather than ``hopeful,''
transfer any extra votes for that candidate
beyond $Q$ to next-choice candidates,
and then proceed to the next iteration without eliminating any candidate.
STV eliminates candidates (transferring all of their votes as in IRV)
only when it reaches a stage
where {\em no} hopeful candidate has yet reached the quota $Q$
but there are still more than $n$ uneliminated candidates.

\paragraph{The Difficulty of Transferring ``Extra'' Ballots from Elected Candidates:}
This refinement creates a significant second-order difficulty, however:
since only {\em some} votes for elected candidates are to be transferred,
{\em which} specific ballots are to be transferred?
This question matters, and can affect the election's outcome
for the remaining candidates,
because each ballot may have a different rank-ordering.
For example, suppose the election authority takes the extra ballots
from the ``top of the pile,'' those tend to be ballots cast most recently,
and A's late-voting supporters tend to rank their ballots ABC
whereas a similar number of A's earlier-voting supporters tend to vote ACB.
Then candidate B may have a significant advantage over C
and benefit from many ABC ballots being transferred
and most ACB ballots left behind.
\xxx{ mention/document this as a real concern in Cambridge STV elections?} 

As a result, many STV systems require
that the choice of the ballots to be transferred be random.
Introducing randomness into the vote-counting process
brings further problems, however:
it is harder (and perhaps impossible) to perform a precise recount
without the outcome varying purely due to a change in the random choices;
it is hard to verify independently that the randomness used 
was {\em truly} random and unbiased; \xxx{cite}
and in general voters may rightfully question whether and why
part of the election process should be run like a lottery.

Here again, taking the perspective of voting power as a liquid may be helpful,
this time in a way that leverages
both the {\em divisibility} and {\em flow} properties of physical liquids.
Some STV variants,
such as Meek's~\cite{meek69nouvelle,hill87algorithm}
and Warren's~\cite{warren94counting},
rather than transferring a selected fraction of (whole) ballots
from elected candidates,
instead transfer {\em all} of the ballots from elected candidates
but at a {\em fraction} of their original voting power.
This refinement allows STV to operate with {\em no} randomness
except in the typically-rare case of perfect ties --
but the mechanisms are complex, subtle,
difficult to explain to ordinary voters,
and even the experts have trouble agreeing
on the precise rules~\cite{hill05meek}.

\begin{figure*}[t]
\centering
\hfill
\subfloat[First round: voting power from all ballots
	flows to each voter's first-choice candidate.
	Popular candidate $A$ with four first-choice votes
	exceeds the 3-vote quota needed to win a seat.
]{\includegraphics[width=0.45\textwidth]{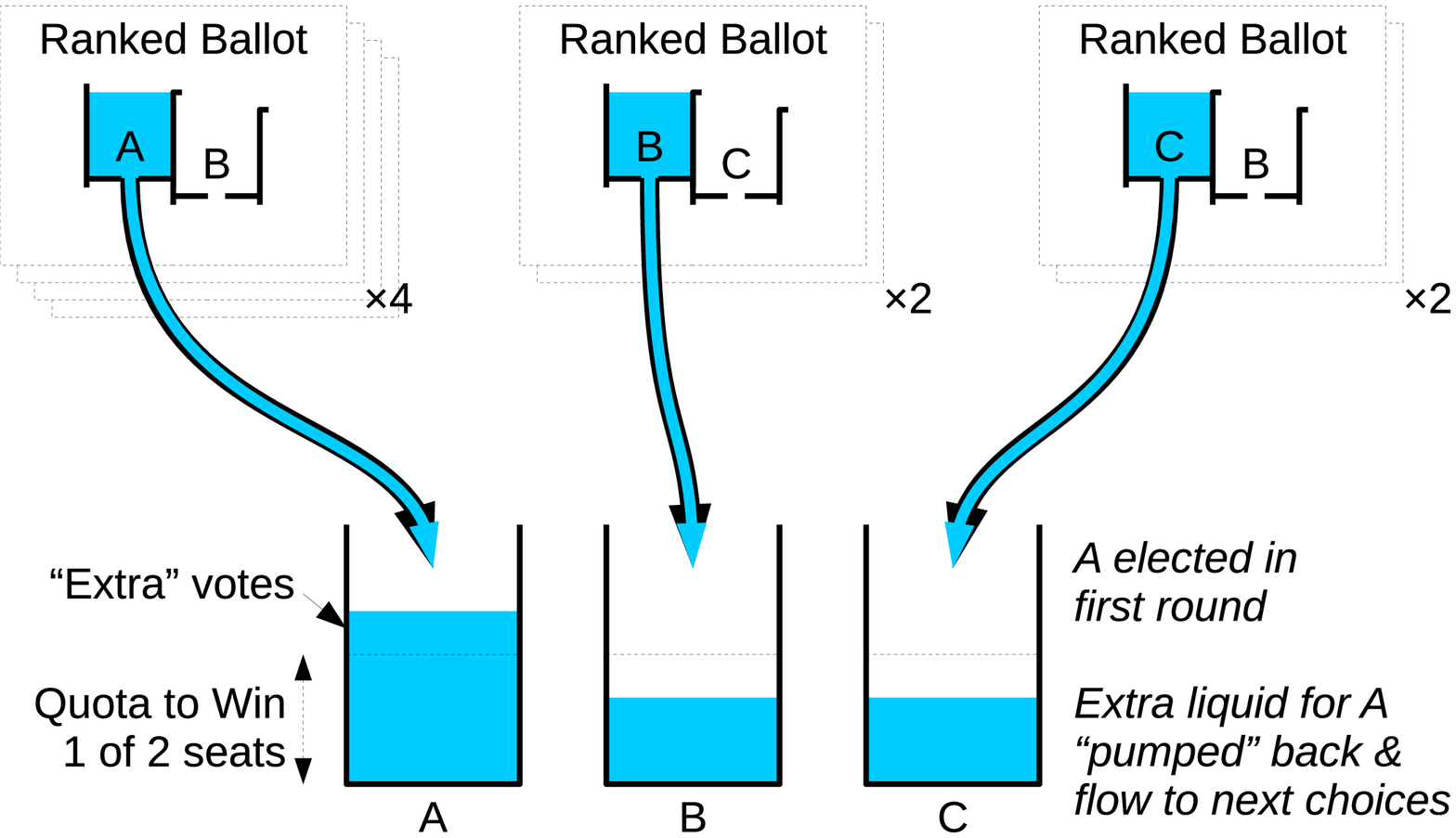}}
\hfill
\subfloat[Second round: the extra $1/4$ of $A$'s voting power
	is ``pumped'' back to refill $A$-voter ballots {\em partially},
	allowing their second-choice preferences to help elect $B$.
]{\includegraphics[width=0.45\textwidth]{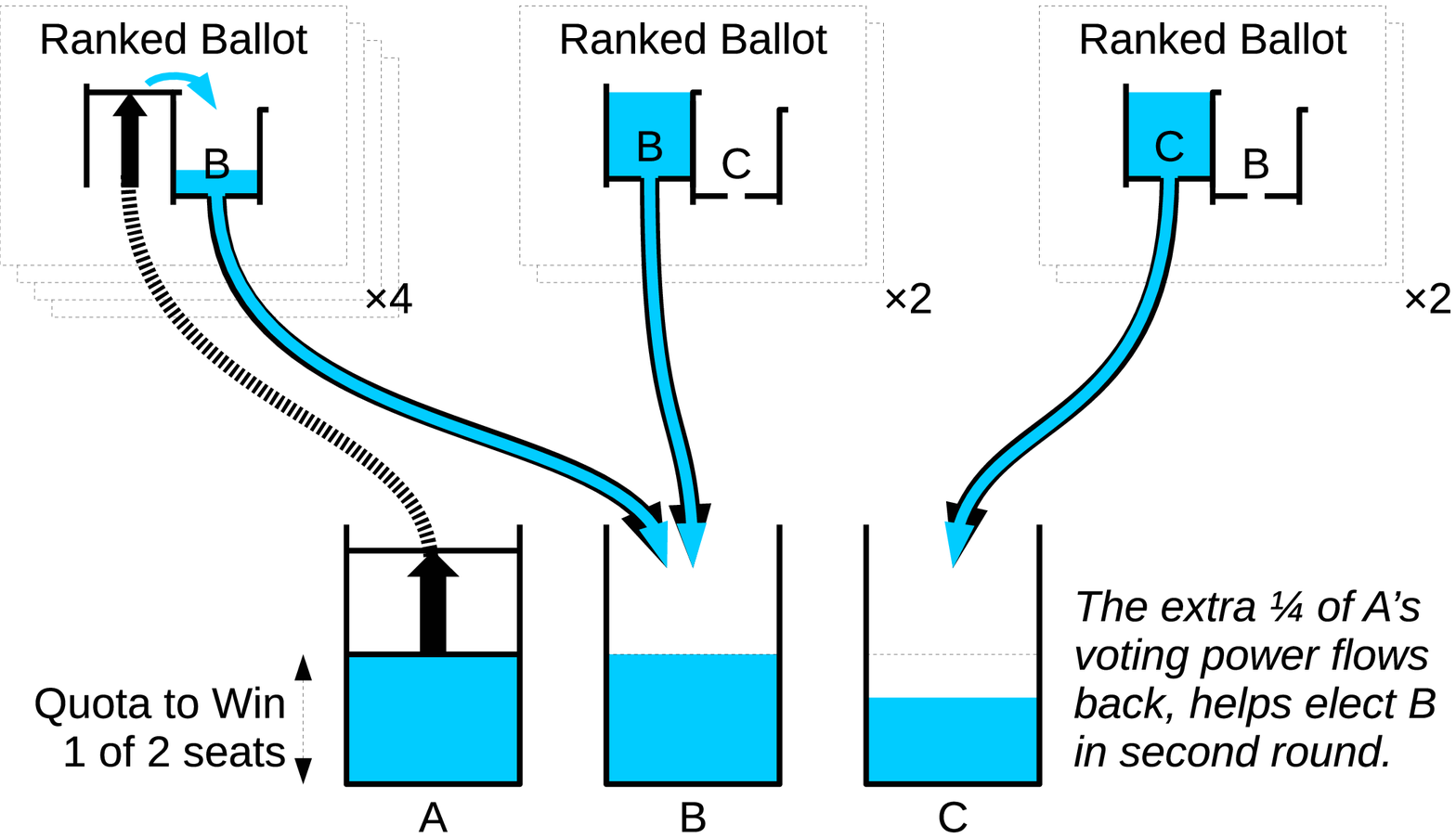}}
\hfill
\caption{Single Transferable Vote (STV) example illustrated
	via subdivisible liquid flows of voting power}
\label{fig:liquid:stv}
\end{figure*}

While the liquid analogy by no means makes STV ``simple,''
nevertheless it may be helpful in intuitively understanding, and explaining,
the fractional ballot-transfer in STV variants like Meek's and Warren's.
Consider the example in Figure~\ref{fig:liquid:stv}
of an STV election with eight ballots, three candidates, and two seats.
In the first round, popular candidate A easily wins one seat,
accumulating voting liquid from four ballots
despite needing only three to win.
To avoid wasting this extra voting liquid,
A retains only the 3-vote quota needed to win a seat,
and the extra voting liquid is ``pumped'' back
and distributed evenly among the ballots that had voted for A,
only {\em partially} refilling each ballot.
This excess voting power of A's voters then flows
to each ballot's next-choice candidate, in this case B,
who then has enough aggregate liquid to win the second seat.

The liquid analogy may thus help clarify why it makes sense
for Meek's or Warren's versions of STV to transfer fractions of votes.
In this analogy,
all three interesting properties of liquids are relevant:
the fact that liquid flows
(voting power is both delegable to candidates
and transferable to next choices),
that liquid is subdivisible
(an elected candidate retains only the required fraction
of voting power they received),
and that liquid is volume-preserving
(each voter's equal amount of voting liquid always ``goes somewhere''
and is neither lost nor artificially expanded,
and most of it goes toward electing the $n$ winning candidates).
In these ways, the mature versions of STV seem to represent
a strong, established precedent for key elements of liquid democracy.

Of course, the liquid analogy is not perfect,
and in particular does not precisely reflect some of the subtleties
of either Meek's or Warren's systems --
but those subtleties tend to be corner-cases
likely to affect outcomes extremely rarely in practice.
Further, the liquid democracy perspective might in fact
suggest further potentially-interesting tweaks to STV,
such as eliminating the need to calculate the winning quota
as an {\em integral} number of votes as in the Droop quota:
the ``floor'' and final $+1$ parts of the quota calculation
are probably unnecessary since a non-integral ``amount of voting liquid''
is not nearly the problem
that a non-integral ``number of votes'' may sound like.
\xxx{ cite Hagenbach-Bischoff quota }

\subsection{Risks and Disadvantages of Ranked Voting Systems}

While IRV and STV hold considerable appeal,
they also have well-known disadvantages.
Asking voters to rank the candidates is arguably more complex
than simply asking voters to pick one, 
or even to mark whichever ones they support as in Approval voting.
Paper IRV and STV ballots with many candidates
can require large $N \times N$ square matrices of ovals to fill,
scaling poorly to elections with many candidates.

In addition,
IRV and STV are often criticized as tending to prefer extremists over moderates,
because middle-of-the-road candidates who may be many voters' second choice
but not many voters' first choice will be eliminated early,
leaving a contest between more extreme candidates
with focused bases of first-choice support.
To avoid early elimination,
centrist candidates must defend themselves ``on both sides''
from more extreme competitors,
while the extremists might need to defend only one front --
and if the first-choice voter support for centrist candidates is split,
then all of them may well be eliminated
before transferred votes start having an impact.
For this reason,
an informal ``rule of thumb'' in IRV/STV campaigning
is that only first-choice votes matter:
if you can't convince the voter to make you their first choice, move on.
It is thus questionable whether IRV or STV truly enable voters
to avoid strategic thinking and ``vote their conscience''
by supporting a first-choice candidate who has little chance of winning.

Besides the risk of overloading voters with ``too much choice,''
which we will address further below,
the information-richness of ranked voting creates
a more subtle risk of coercion and vote-buying.
Since the mature STV implementations
typically require computers to perform
the nontrivial ``flow'' calculations,
for transparency it is desirable for the ``raw'' list
of anonymized ballots to be published in full
so that anyone can independently repeat and verify the STV calculations.
However, even if all IRV ballots are fully anonymized
and disconnected from their voters' identities,
either physically (via shuffling in a ballot box)
or electronically (via verifiable cryptographic shuffle),
each ballot's rank-choice list carries enough information
to make ballots potentially uniquely-identifiable.

An STV ballot asking voters to rank only 10 candidates, for example,
offers 3.6 million (10 factorial) possible complete rankings.
	\xxx{ largest STV ballot? some evidence: \url{https://en.wikipedia.org/wiki/Ranked_voting} }
A vote-buyer might ask a voter to cast an STV ballot
with a particular ranking of all 10 candidates,
the first one or two of which the vote-buyer chooses
according to his preferences and the remaining ones randomly,
which will effectively ``watermark'' the requested ballot 
by making it unique with high probability.
The vote-buyer then simply watches for this specific ranking to appear
in the published election outcome,
and pays the voter only if it does.

\subsection{Cumulative Transferable Vote (CTV):
		Proportional Representation with Vote Spreading}

\xxx{ illustrate? }

Viewing existing voting systems such as those above
through the lens and framework of the liquid voting power analogy,
for purposes of both vote spreading (approval, cumulative, quadratic voting)
and vote transfer (IRV, STV),
makes other potentially-interesting variations more readily-apparent.
Observe that IRV and STV maintain a ``winner-take-all'' approach
to interpreting ballots {\em in each round},
in that each voter's ballot can help only ``one candidate at a time'';
vote transfer merely allows a ballot
to help different candidates in different rounds.
There appears to be nothing fundamentally essential about this
``serial winner-take-all'' approach, however.
We could readily combine the vote-spreading mechanisms
of approval, cumulative, or quadratic voting
with the transfer mechanisms of IRV and STV,
and there may be interesting advantages to doing so.

First consider a variant we'll call {\em cumulative transferable vote} or CTV,
a combination of cumulative voting we propose with STV-like vote transfer
for multiwinner elections.
In this variation, voters conceptually divide
a fixed amount of ``voting liquid'' into parts
and assign them to candidates as they prefer,
exactly as in Figure~\ref{fig:liquid:cum}.
Elections could also use the paper ballot approach
in Figure~\ref{fig:liquid:cum-paper}
with multiple fillable ovals per candidate,
in which voters may fill any set of ovals
to divide and assign their voting power in that many parts.

Given a set of cast ballots expressed this way,
however, we use a multi-round process as in STV
to tally them and choose winners.
In each round, we first check if any candidates have more than
a winning quota of $1/(k+1)$ fraction of the total voting liquid in play,
where $k$ is the number of candidates not yet elected or eliminated.
If so, each candidate above this threshold is marked elected,
and exactly as in Meek's STV~\cite{meek69nouvelle,hill87algorithm},
any surplus voting liquid above this threshold
is returned to the supporting voters' ballots
in proportion to the voters' respective contributions.
That surplus voting liquid is then redistributed to any other candidates
still in play (neither elected nor eliminated)
on the respective voters' ballots,
while preserving the voter-specified allocation ratios
among those remaining candidates in play.
If in some round there is no candidate over the $1/(k+1)$ threshold,
then the candidate with the least support (voting liquid) is eliminated,
exactly as in STV,
and {\em all} the liquid that went to supporting this candidate
is returned to the supporters' ballots for proportional redistribution
among other candidates still in play as above.

This variation has several attractive features.
It assures proportional representation while minimizing ``wasted votes''
and reducing strategic voting incentives as in STV, 
with a ballot structure nearly as simple as in plurality or approval voting --
or even just as simple,
if we reduce it to a ``one fillable oval per candidate'' ballot.
By allowing voters to assign multiple parts to candidates
as in Figure~\ref{fig:liquid:cum-paper},
we give voters the ability to express both equal and unequal
amounts of support to different candidates,
whereas IRV and STV only meaningfully allow voters
to express unequal levels of support through ranking.
Thus, CTV combines the ballot simplicity of approval voting,
the richness of expression of expression of cumulative voting,
and the proportional representation properties of STV.

Another potentially appealing property of CTV is that
voters who split their support among multiple candidates
will be helping {\em all} of them (at least some) in early rounds,
which may help moderate candidates with broad but diffuse support
avoid early elimination before vote transfer can start helping them.
Consider for example a voter who splits his vote at a 2:1 ratio (3 parts total)
between most-preferred specialty candidate A who is unlikely to win
and a more mainstream candidate B with broader but less-focused support.
In CTV, even though specialty candidate A gets more of the voter's power
in the initial round, mainstream candidate B gets some of it from the start,
reducing B's risk of early elimination if many voters spread their vote
between different specialty candidates and mainstream candidate B.
If specialty candidate A is eliminated,
then {\em all} of the example voter's power
is transferred toward helping B in subsequent rounds.
And if either A or B is elected before the other is eliminated,
then any excess voting power is transferred to helping the other.

A third appeal in CTV is the simplicity of its vote-tallying calculations
in relation to state-of-the-art STV variants.
In Meek's STV, for example,
the vote transfer process for each round is a complex calculation
that itself requires an iterative successive approximation process
amounting to solving a Linear Programming (LP) problem~\cite{vanderbei13linear}.
CTV, in contrast, has no need
either for random selections of ballots to transfer
or for iterative approximations;
all the relevant voting power or ``liquid transfer'' adjustments
can be done with a single straightforward pass through the ballots
in each candidate election/elimination round.

These appealing characteristics of CTV
are only informal and intuitive, of course.
We make no pretense of being able to perform a thorough analysis of CTV here;
such an analysis and comparison with other alternatives
remains for future work.

\xxx{
make sure the level of expressiveness is clear somewhere:
maximizes expression in individual vote.
Voter can vote for a single option;
can spread the vote evenly over several options as in Approval voting;
can express relative preferences between different options as in ranked systems;
and can express the degree of those differences in preference.
}

\subsection{Quadratic Transferable Vote (QTV): Rewarding Vote Spreading}
\label{sec:liquid:choice:trans:quad}

\xxx{ illustrate? }

A similar combination of the liquid elements of quadratic voting and STV
suggest another intriguing hybrid
we'll call {\em quadratic transferable vote} or QTV.
In this variant,
we give each voter an equal measure of virtual liquid,
which they can divide into parts and use to fill virtual ``water balloons''
for the candidates they support,
as illustrated in Figure~\ref{fig:liquid:quad}.
The simplified paper ballot structure of Figure~\ref{fig:liquid:cum-paper}
is also possible here,
since QTV identical to CTV in the way voters express preferences
but differs only in the way the results are calculated.

As in STV and CTV, we calculate the results in multiple rounds,
electing candidates whose stack of virtual water balloons
exceeds the relevant $1/(k+1)$ height threshold
when $k$ candidates are in play,
or eliminating the weakest candidate if no candidate is above the threshold.
When a candidate is elected,
we ``deflate'' all the balloons in the winning candidate's stack
so that their aggregate height exactly matches the threshold.
In this deflation process,
we preserve the proportions
among the balloons' respective heights (diameters),
\ie, preserving the percentage of total height
each voter contributed toward electing the candidate.
The liquid we recover from deflating the elected candidates' balloons this way
returns to the supporting voters' ballots,
just as in CTV,
for redistribution to other candidates still in play
according to the user's expressed preferences.

As in quadratic voting,
QTV effectively rewards voters for spreading their vote --
especially in early rounds when many candidates remain in play --
because the aggregate stack-height impact of a vote spread widely
is greater than the stack-height impact of a vote plumped onto one balloon.
This property effectively encourages voters
to allocate at least one part of their vote
to each candidate they support at all,
while allowing them to ``pay more'' (at an efficiency cost)
to help their most-preferred candidates more than others.
We may also expect QTV to help candidates with a broad base of support,
even if many of their supporters spread their vote among several candidates,
perhaps further reducing the risk of early elimination
to moderates with a broad but diffuse support base.

If a voter decides to support two candidates equally instead of just one,
for example,
then each of those candidates still individually get $\sqrt{1/2}$,
or about 70\%,
of the ``help'' they would get if the voter supported them alone.
A voter who splits his vote equally among four candidates
helps each one 50\% as much as he would by plumping on one alone,
and thus helps the four candidates $2\times$ as much in aggregate.
Because a coalition of candidates receives more aggregate help
from voters who spread support among all of them,
this effect may incentivize candidates to ``be civil,''
build or join coalitions, and encourage voters to support
other candidates they consider reasonable in addition to themselves.

A voter who strongly prefers specialty candidate A
but also supports more mainstream candidate B at a 3:1 ratio
helps A 82\% as much as by plumping,
while also helping B 50\% as much.
If the voter's preferred candidate A is eliminated,
then all the voter's liquid transfers to support B at 100\% rather than 50\%.
If A is elected,
then only the excess voting liquid that ``wasn't needed'' to help A
gets transfered to B,
subsequently supporting B at some level between 50\% and 100\%
of a plumping vote.

In summary, QTV appears attractive in that it
allows and encourages users to support multiple reasonable candidates,
enables users to express both equal and unequal strength of preference,
ensures proportional representation and avoids wasting votes as in STV,
and like CTV is much simpler to calculate than non-random STV variants
such as Meek's.
To whatever (perhaps limited) extent
that real voters match the ideal rational model that quadratic voting assumes,
QTV should incentive voters to express their ``true strength of preference''
in deciding which candidates to support and by how much.
These features suggest QTV is promising,
but we leave many questions and subtleties for future work,
such as precise analysis of QTV's properties in comparison with other systems,
and the handling of subtleties such as the negative votes
that quadratic voting proposes to allow.

\xxx{
\subsection{Generalization to Vector Transferable Vote (VTV)}
\label{sec:liquid:vector}

p-norms, explore general properties, ...
}

%% file: liquid/deleg.tex
\section{Liquidity in Delegation to Simplify or Aid in Choice}
\label{sec:liquid:choice:deleg}


The above sections have discussed precedents that increase
the richness of voter choice,
to allow splitting vote power and expressing (relative) strength of preference
(Section~\ref{sec:liquid:choice:spread}),
to avoid ``wasting'' votes in majoritarian or proportional elections
(Section~\ref{sec:liquid:choice:trans}).
Increased richness of choice also place a burden on voters, however,
many of whom may have little time or inclination
to study carefully and come to an informed understanding
of the choices available.

For this reason, many voting and ballot schemes make provisions
to simplify, manage, and perhaps even {\em decrease}
the richness of choice available,
often by enabling (or even requiring) voters to {\em delegate}
many details of their choices,
most commonly to their preferred party.
Taking the liquid analogy,
we can view such delegation as {\em transitive} flows of voting power:
\eg, from voters to parties,
then from parties to candidates,
and finally from elected representatives to decisions
on specific laws or issues.

\subsection{Precedents for Delegation to Simplify and Manage Choice}

We observe two clear precedents for delegation of democratic power
as a way to simplify or manage choice:
first, the basic structure of representative democracy,
and second, political parties.

\subsubsection{Representative Democracy:}

Small communities sometimes practice {\em direct democracy},
in which ideally all eligible voters discuss, deliberate on,
and participate in decions on all significant decisions
affecting the community.
Because the complexity of governance
and the total number of decisions to be made 
tends to grow proportionally with the size of community, however,
the pure ideal of direct democracy rarely functions well
beyond small communities of tens or at most hundreds of participants.
This scaling challenge,
caused by the inevitably limited time and attention of ordinary citizens,
necessitated the now far-more-common representative form of democracy,
in which voters only periodically elect representatives who specialize
in carrying out the day-to-day tasks of political decision-making.

Although pure direct democracy does not readily scale to large democracies,
elements of direct democracy such as popular initiatives and referenda
became a fixture of the Swiss federal constitution in the 19th century,
and have since been adopted in many governments around the world.
In this hybrid approach,
while elected representatives still handle most decision-making,
the electorate is directly involved in select decisions on major issues.
In effect, the people delegate their political power
for {\em most} day-to-day governance decisions to their representatives,
but retain direct involvement in certain decisions,
including the potential ability to override decisions of their representatives.
This general notion of {\em power delegation with a possibility of override}
is a recurring idea we also see in party systems, discussed next,
and generalized further in today's liquid democracy ideas,
which we will return to later. \xxx{forward ref}

\subsubsection{Political Parties and Straight-Ticket or Party-List Voting:}
\label{sec:liquid:choice:straight}

While political parties in practice serve many functions,
one of them is to simplify the choices of voters who may not have
sufficient time or interest to keep up
with the many candidates or issues on a ballot individually,
but feel a close-enough affiliation with some political party
to leave many of the details of their decisions to their party:
\eg, by voting for candidates mainly because of their party affiliation.

In most governmental elections it is typical for ballots
to display party affiliation prominently alongside candidates' names,
and making it easy and fairly common (though by no means universal)
for voters simply to vote for all the candidates from their preferred party
if they choose.
Certainly it is common in many elections for ballots to list
many candidates across multiple open positions,
about whom many voters are likely to know little to nothing.
Thus, many of the votes that go toward electing ``down-ballot'' candidates
are likely to be party-line votes,
whose candidates may benefit substantially from the ``coattail effect''
from voters showing up mainly to vote for some other
more prominent candidate or issue. \xxx{cites}

Historical US elections even used ballot designs
in which each each party had its own differently-colored ballot,
making voting easy for straight-ticket voters
but more difficult and confusing for split-ticket voters,
who would need to mark and cast two or more
differently-colored ballots. \xxx{cite}
Even today, a number of states use ballot designs
allowing voters to cast a straight-ticket ballot
for all their party's candidates by marking a single option. \xxx{cite}
Texas not only provides this option,
but also allows voters to {\em override} the party-line vote
for specific races by marking a candidate from another party
for that race. \xxx{cite}


Many European election systems take this philosophy
of simplifying choice through party delegation even further,
in the form of party-list proportional representation.\xxx{cite}
In such designs, voters choose {\em only} parties,
each of which has published a ranked list of that party's candidates
for the available seats.
After ballots are counted,
each party that reaches a minimum popular vote threshold
is awarded a number of seats in proportion to the total vote for that party,
in the ranked order defined by the party.
Party-list proportional representation can thus be viewed
as a constrained variant of STV,
in which voters cannot specify their own candidate ranking
but must instead choose among a small number of predefined rankings,
one per party.

\xxx{ liquid illustration }

\xxx{ discuss threshold considerations and anti-extremism?
	propose vote-splitting with quadratic voting
	as a potentially interesting alternatve approach?
	(or maybe better later since that's not really a precedent. }

\xxx{ Subsection on voting-advice cards, systems, platforms?
	Observe that when these are used, their influence on decisions
	is effectively hidden from the electoral system and most polls;
	they have to pass no evaluations or tests in order to be used;
	no one is necessarily accountable if such tools were to be found
	to be deliberately misleading voters since it's just ``speech.''
}

In summary, allowing -- or even requiring -- voters to delegate many or all
specific details of their choice to their preferred party
is a common and accepted practice,
if not without many potential strengths and weaknesses.\xxx{cite}

\subsection{Few Versus Many Parties:
		Countering or Merely Obscuring Extremes?}

Political parties have become a ubiquitous structural mechanism
not only to organize races for power in democracies,
but also to simplify voters' choices
by enabling (and sometimes requiring) them to delegate
details of their decisions to their preferred party.
The {\em number of parties} people have to choose from, however,
is usually severely limited for structural reasons.
Two-party systems such as the US based on ``winner-take-all'' elections
strongly discourage the rise of third parties due in part to fear of the
spoiler effect~\cite{allen05roots,rapoport05threes}.
While the proportional representation systems common in Europe
more readily accommodate multiple parties,
to be politically relevant they must have enough direct support
to obtain at least one representative seat.
Further, to obtain any representation
parties must often pass a legal threshold (\eg, 5\%),
typically imposed out of fear of extremist parties~\cite{carter02proportional}.


These structural constraints favoring large parties
force them to aggregate and effectively {\em hide}
the complex organizational and activist structures within parties,
the issue-centric campaigns and influence structures outside of parties,
and the vast constellation of actual reasons people cast the votes they do.
How many people voted for a particular party
due to general alignment with the party's platform or ideology,
how many due to alignment on
a {\em single} issue the voter considers most critical,
how many due to personal attraction to a particular candidate,
and how many solely because the opposing party or candidate seems worse?
After an effective local activist or issue-focused organization
successfully persuades a citizen to town out and vote for their cause,
the electoral system obscures that fact in an anonymous statistical bump,
aggregated with all the other distinct reasons that other voters
turned out (or didn't) to vote for particular candidates or issues.
This aggregation leaves the {\em actual} causes of influence to be
merely guessed at by pundits and journalists,
or estimated via bias- and error-prone
opinion and exit polls~\cite{curtice08exit,ball18donald}.

\xxx{
You no longer need pollsters to attempt the difficult task of
getting good information about {\em why} people voted YES or NO
against statistical error
and significant bias issues\cite{curtice08exit,ball18donald}.
}

Limiting choice obscures not only the motivations of voters
but also the motivations of {\em non-voters}:
the often-sizeable percentage of the eligible electorate
who do not actually vote.
For example,
does the silence of a non-voter express lack of interest in general,
lack of time to go vote or meet associated
registration or identity process requirements, \xxx{cite disenfranchisement}
or does silence represent a vote for ``none of the above''---merely
a disapproval of those few particular choices that were laid before the voter?
How many of today's non-voters might become voters
if they were given the choices they actually desire?
While online digital forums can obscure the motivations
of silent communities~\cite{gangadharan18digital,ananny18silence},
offline political structures set far earlier precedent for such exclusion.

We have many reasons to suspect that structural limitations
on the breadth of party choice may not so much {\em prevent} extremism
so much as sweep it into a corner and temporarily out of sight,
until it suddenly bursts through or circumvents the structural barriers
intended to contain it~\cite{carter02proportional,golder16far}.
Even if the US's two-party structure
prevented the Tea Party from technically being a true political ``party,''
it did not prevent the movement
from fundamentally shifting the Republican party's
positions and discourse~\cite{abramowitz12grand,skocpol13tea}.

When partisan competition gives rise to cultural polarization and tribalism,
identity politics can lead people to vote
against their own interests~\cite{taub17why}.
When the working structures of political organization
must reach large population segments to win votes,
they become dependent on money to achieve those scales,
incentivizing representatives to exploit cultural divisions and fears
to win votes while quietly sculpting their policy choices
around the interests of their elite
donors~\cite{gilens14theories,smith14political,speck13money}.

\xxx{ two-party systems incentivize political leaders
to keep their citizens in the dark.
From Henry Milner "Civic Literacy: How Informed Citizens Make Democracy Work":

In comparison to FPTP, under PR there is less incentive for political leaders,
who may very well need their opponents’ support after the election, to inhibit
the awareness of the electorate of alternative positions on the issues of the
day. In contrast, (the) FPTP-based ... governing party is expected to implement
its program as if a majority of the population was behind it, and not seek
broad-based multipartisan support for needed, but controversial, reforms. It
knows, moreover, that such support is rarely forthcoming... With politics as a
ruthless zero-sum game... distorting the opponent’s position... while keeping
one’s policies vague, pays off... The result, whatever the intentions of the
individuals involved, is a public less informed than it needs to be. (p. 85)

}

In short, large-scale political organization inevitably relies on
vast networks of fine-grainted structures
both {\em within} and {\em around} the few major parties in a country,
and leaving those structures outside of the electoral system
and their effects hidden in aggregate election results
both deprives us of important information
and exposes those structures to corruption.
Can we, and perhaps should we,
bring these fine-grained structures into the sunlight
and make them both visible and accountable?


\xxx{ 
Hiding these structures may expose them to corruption.
The tax plan ``Con Job''
}

\xxx{ Should do a better job at referencing and talking about
	Farrell and Schartzberg's paper on epistemic authority,
	and the way people intellectually delegate to those authorities.
}

\subsection{Revealing Choice Structures through Transparent Delegation}

\xxx{ one perspective: giving advocacy groups official status.
	separate from party-as-platform/ideology notion? }

\xxx{ clarify the issue of transparency in debate in the public sphere
	with respect to ~\cite{landemore18democracy}}

\xxx{ discuss somewhere how the lack of transparency in the {\em why}
voters vote vor or against something gives maneuvering room and cover to elites
to manipulate information and interpretations of voter preferences:
see for example \cite{hacker05abandoning} on tax cuts,
where surface questions suggest that voters were in favor of tax cuts,
but didn't take into account deeper priorities or considerations...
elites get to claim voters want something that they don't, or vice versa.
Transparent delegation gives us a tool to unbundle voters' preferences.
}

Since accepted democratic structures already embody
multiple transitive levels of decision delegation --
\eg, from voters to parties to representatives --
it is not such a stretch to envision allowing further levels of delegation,
allowing voters to express their preferences
in terms of more fine-grained organizational structures.
Instead of parties and partisan voters appearing to the electoral system
as a monolithic mass of undifferentiated votes as they do now,
major parties might then represent masses of voting power
collected and aggregated from many smaller organizational structures
by delegation more explicitly and transparently via the electoral system.

Giving voters the power to delegate their choices
is central to many conceptions of liquid democracy.
But if we were to embrace and expand this power of delegation
beyond a choice between a few large parties,
what -- or who -- should we allow voters delegate {\em to}?
We explore four potential alternatives:
delegation to finer-grained parties or {\em microparties},
delegation to issue-centric organizations,
delegation to individuals,
and delegation to algorithms.

\subsection{Delegation to Microparties:
		Expanded Choice among Political Parties}
\label{sec:liquid:choice:microparties}

One way to allow for more fine-grained structures might be
to embrace the concept of {\em microparties}.
We informally define a microparty as an organization
representing a group of voters
that may be too small and narrowly-focused to have a realistic chance
of obtaining even one seat in a large-scale public election.
A microparty might instead, for example,
wield a (smaller amount of) power by 
selectively delegating its supporters' votes to,
and thereby influencing, larger parties.
Microparties might naturally specialize to narrower demographic audiences,
smaller geographic regions,
or more specific party platforms or ideologies than a major party can.
\com{
While major parties generaly have platforms defining positions on many issues,
microparties might be organized largely around a single issue,
a particular population segment, a small geographic region,
or even a particular personality.
}
Microparties along some of these lines have formed spontaneously
in Australia in recent years, in fact,
though as more of an unintended and not-always-welcome side-effect
of unrelated electoral reform rather than by design~\cite{kelly16party},
and thus not necessarily taking a form that we might prefer by design.

\xxx{ To read in more detail: Party Rules?
}

An essential attraction of microparties is that they can
be ``closer'' to their relevant segment of voters,
both in terms of responsiveness to their voters' interests
and in terms of assisting and guiding the choices of their supporting voters.
By giving formal and explicit ``recognition'' to microparties
in the electoral system,
we might not only expand the breadth of party choice voters have,
but also make it an explicit and transparent part of the public record
when they delegate their choices to a microparty.

Suppose, for example, that a microparty
cannot realistically find many candidates of its own
to run for national political offices,
let alone hope to win those races.
However, the microparty can nevertheless campaign among its target population,
and explicitly delegate the votes it receives
to major-party candidates who are most closely aligned to the microparty
on the issues most important to the microparty.
In a Party List Proportional Representation system,
this might in principle be achieved simply by permitting microparties
to publish lists that include candidates from {\em other} parties
when they are not running enough -- or perhaps any -- candidates of their own.

In a Single Transferable Vote system,
microparties can in effect do this in any case,
simply by handing out ``voter advice'' cards to their supporters
showing the microparty's recommended rank-ordering
of the available candidates regardless of party affiliation.
(It may be little coincidence that microparties in some form
emerged spontaneously in Australia, which depends heavily on STV.)
In a multi-election ballot for multiple races for local and regional offices,
a microparty could provide its supporters
a recommended selection of down-ballot candidates for all those offices,
cherry-picking from its own and other parties' candidates
according to the microparty's judgment of best alignment with its position.

\subsubsection{Election System Design for Delegation Transparency:}
With appropriate ballot design or electronic voting user interfaces,
microparty supporters might be able to adopt the microparty's choices
in ``straight ticket'' fashion
or selectively override particular choices,
just as Texas ballots already permit
in straight-ticket voting for major parties
(see Section~\ref{sec:liquid:choice:straight}).
If we allow for a large number of microparties in the interest of choice,
we should not expect -- or promise --
that all microparties would be listed explicitly
on limited-size printed ballots.

E-voting environments can easily solve the ballot scalability problem, however,
by allowing voters to look up the microparty
by the first few letters of its name,
or by scanning a QR code printed on the microparty's printed voter-advice card.
The net effect on the actual electoral outcome would be identical to that
if the voter simply brought along the microparty's voter-advice card
to the ballot box and entered manually entered the recommended choices
(overriding them as desired),
which voters can do anyway in most any election system.
However, besides offering the voter the convenience of automation,
we might hope to ensure that the {\em fact} that a specific number of voters
followed the microparty's recommendations would be tallied
and appear in the election's official outcome,
giving everyone clear information on each microparty's influence structure:
both how many people are following its recommendations
and where (\eg, to which major-party candidates) its influence is flowing to.

If it were not only accepted but expected
that microparties wield voting power mainly by delegating it transitively
to other parties large enough to win seats,
then microparties could use the official, public information about their power
to negotiate and form alliances with larger parties,
the microparty offering formal support for some of the major party's
candidates or issues,
in exchange for the major party's consideration of the microparty's interests.
This type of inter-party dance is already standard procedure
between larger and smaller parties
in the coalition-building stages of multi-party governments.
Delegation through microparties might merely give large and medium-size parties
more choice of ``dance partners'' of more wide-ranging sizes.
Some microparties might form stable, long-term, exclusive associations
with larger parties, effectively becoming subsidiaries,
while other microparties might retain greater independence,
switching alliances regularly or choosing major-party candidates to support
on purely a basis of alignment with individual candidates' positions.

\xxx{ diagram? }

\subsection{Delegation to Single-Issue Organizations --
		Exclusively or Jointly via Vote Spreading}

\xxx{ 
Discuss Robert Dahl's notion of pluralism focused on interest groups.
References to read from hacker05abandoning:
10 For a more detailed survey and analysis of the Old and New Pluralism, see Pierson 2001.
11 Campbell et al. 1960; Converse 1964.
13 Lindblom 1977; Olson 1965
and also the "new pluralism"...
"Effectively managing information and employing cogni- tive cues and shortcuts, voters can produce outcomes not so different from those of fully informed individuals.15"
(but to what extent are they doing so through herd mentality,
by relying on {\em other} hopefully-more-informed leaders to think for them?
and can we even know how much and how -- can we make this transparent?)

"Voters often lack important resources needed to exercise effective control over policy. Perhaps the most crucial resource is knowledge-knowledge of what politicians do, and of what their actions mean for
citizens."
}

\xxx{ link to ~\cite{lee18societal}
and their look at crowdsourced activism organizations like MoveOn and GetUp!  
To whatever extent such methods are effective,
how can we know how many people successfully influenced an election
{\em because of} their involvement in such issue-focused mechanisms?
}

As mentioned above, one ``limit case'' of a microparty is one
that that might make no pretence of representing an entire ideology or platform,
but instead is organized around a single issue or cause:
\eg, a particular position on health care, environment, human rights, jobs, etc.
The existence of such single-issue microparties
might seem slightly absurd and dysfunctional 
in an election system that allows voters to choose
only {\em one} candidate or party for a particular office:
how could we reasonably expect anyone to choose an organization
representing only {\em one} issue to represent them and guide their choices?

One reason that allowing single-issue microparties might be desirable, however,
is to reveal the prevalence (or lack thereof) of ``single-issue voting.''
If a certain segment of voters cares about one issue above all else,
and decides to support this microparty
(and, by delegation, the candidates it recommends)
in preference to all the broader platform-based parties or microparties,
then it may be useful for the major parties, and the general public,
to know this: \ie, to understand the perceived importance of that issue.
We cannot in any case prevent voters from choosing a particular candidate
because of their stance on a single issue,
but we can hope to gain understanding of
how often that happens and on which issues.
Allowing voters to delegate through issue-based microparties would enable that.

As we explored earlier in Section~\ref{sec:liquid:choice:spread}, however,
there is no fundamental reason we must force voters
to make {\em only one} choice in an election,
and the same is true for delegation.
Suppose we design an election system allowing voters
not only to delegate their vote but also to subdivide and spread it
among multiple microparties they support,
perhaps with the ability to express strength of preference as well
as with cumulative (\ref{sec:liquid:choice:spread:cum})
or quadratic voting (\ref{sec:liquid:choice:spread:quad}).
This would effectiveley allow voters the freedom
to ``construct their own platform'' if they choose.

For example, if a voter
strongly supports the position of issue-based microparty A
and weakly supports the positions of issue-based microparties B and C,
she might delegate her vote power to them at a 2:1:1 ratio.
With liquid cumulative voting, for example,
a candidate supported only by party B
will receive $1/4$ of the voter's delegated vote.
A candidate who for whatever reasons is aligned with and supported by
all three of these microparties A, B, and C
will receive all of the voter's delegated vote, but via different paths.
If the electoral system statistically reveals all of these delegations,
then the major parties and the public obtain valuable information about
not only the extent to which different candidates are supported,
but also {\em why} they received support:
\ie, through which delegation paths in which amounts.

Of course, no voter would be obligated to ``build their own platform'' this way:
voters would still be free to delegate
to more conventional parties or microparties
representing platforms bundling many issues together.
And once again, obtaining accurate information through the electoral system
about how many voters choose a bundled platform or ideology,
versus building their own platform through vote-splitting,
might represent a valuable public good.

\subsection{Delegation of Decisions to Individuals}
\label{sec:liquid:choice:individuals}

Delegation of voting power to an arbitrary designated individual
has long been accepted practice in corporate governance structures
allowing proxy voting~\cite{riddick86rules,davis07business,agrawal12corporate}.
The idea of allowing voters to delegate to individuals of their choice
in general political processes has been a recurring idea
suggested by many people including
Heinlein~\cite{heinlein66moon},
Tullock~\cite{tullock67toward},
Miller~\cite{miller69program},
Lanphier~\cite{lanphier95model},
Ford~\cite{ford02delegative},
Sayke~\cite{sayke03liquid,sayke03voting},
Green-Armytage~\cite{green-armytage04proposal,green-armytage05direct,green-armytage14direct},
Alger~\cite{alger06voting},
Boldi et al~\cite{boldi11viscous},
and others.
\xxx{more from my blog post, LiquidFeedback, etc.}.
Taken to its limit,
this approach in essence gives voters complete freedom
to choose individually who will represent them politically,
rather than being constrained to a few official candidates or major parties --
or even to a large but still-limited range of microparties.

Many of these proposals envision electronic deliberation systems
reducing the costs of direct participation in legislative processes enough
that anyone who wants to can serve as a legislator,
participating in legislative debates and committees
while wielding the power of whatever arbitrarily large or small group of voters
delegated their vote to them.
As Alger suggests,
``we could have legislatures that could allow quite large numbers
and still work well,
possibly even allowing individuals that represent
only themselves''~\cite{alger06voting}.

Even with suitable electronic systems, however,
it is not easy to reduce the costs of direct participation to zero.
In any real-time deliberative public forum
there tends to be a limited amount of total ``speaking time''
whether the forum is physical or virtual,
and even in a message-based online forum (\eg, Twitter)
there is a limit budget of total human {\em attention} to messages being sent.
\xxx{ attention economy citations? }
Ford~\cite{ford02delegative} and Alger~\cite{alger06voting}
propose apportioning speaking time and other scarce resources
to delegates in proportion to the delegated votes they wield,
and Alger suggests that this objective ``pecking order''
also be used instead of seniority to decide organizational matters
such as to prioritize selection of legislators to committees.
If despite these provisions it proves necessary to limit direct participation,
both Ford and Alger propose STV-inspired methods of eliminating delegates
with the least support from direct legislative participation,
while still allowing eliminated delegates to influence the decision process
by transitively ``re-delegating'' their voting power (Ford)
or filling out incomplete preference lists of their supporting voters (Alger).

\xxx{ in Shubik's analysis:
liquid democracy presents a way for individuals in the fourth class
to rely on their friends in the third class for
more nuanced but representative decisions.
}

\subsubsection{The Risk of Accidental Dictators:}
Another issue with delegation to individuals
is the potential risk that voting power might become overly concentrated
if many people delegate their vote to a popular celebrity or ideologue.
The German Pirate Party's internal experiments with liquid democracy,
in which one professor happened to accumulate a large amount of voting power,
anecdotally confirm the plausibility of
such risks of creating ``accidental dictators''
through delegation~\cite{becker12liquid}.
Ford's proposal anticipated this risk,
allowing voters to split their voting power
and spread it among multiple delegates~\cite{ford02delegative}.

To mitigate this risk further,
we could use Liquid Quadratic Voting
(Section~\ref{sec:liquid:choice:spread:quad})
to incentivize voters to ``spread out'' their delegated voting power
and actively {\em reward} them for doing so.
For example,
a voter who splits her voting power in equal parts to delegates A and B
would increase each delegate's voting power by $\sqrt{1/2}$,
or 71\% of the amount of power either A or B would receive
if the voter delegated to them alone --
but {\em in combination} the two delegates receive $2\sqrt{1/2}$
or 141\% of the baseline amount of voting power.
Voters who spread their delegated voting power more widely
receive even more of an effective ``bonus'' in aggregate voting power:
\eg, a voter delegates equally to four delegating
helps each of them $1/2$ as much as with single delegation,
but helps the group of them $2\times$ as much in aggregate.
Stated another way,
a celebrity or ideologue whose voters support only her
would need twice the number of total supporters
to wield the same voting power as a coalition of four delegates
whose supporters spread their power evenly among the four.

\subsubsection{The Anonymity Versus Accountability Conundrum:}
Another problem with delegation to individuals
is that we generally want an individual voter's choices
to be private and anonymous,
to mitigate the risk of coercion or vote-buying --
but it seems that delegates who wields the voting power of others
need to vote {\em publicly}
in order to be accountable to their supporters.
If we allow {\em anyone} to become a delegate
with no required threshold of support or other barriers to entry,
then anyone who wants to coerce a voter or buy their vote
can simply require them to become a delegate
(perhaps representing only themselves)
and thus vote publicly,
allowing the coercer or vote-buyer
to verify that the voter ``stayed bought.''~\cite{ford14delegative}.

One way to mitigate this coercion risk
might be to require a delegate to pass a certain threshold of voter support
before being allowed to wield delegated votes and cast votes publicly.
It is not clear how best to set such a support threshold, however:
setting it too high would eliminate the voter freedom and breadth of choice
that makes delegation to individuals attractive in the first place.
On the other hand, it is not clear there is {\em any} support threshold
``high enough'' to make a delegate impervious to coercion or vote-buying,
as suggested by evidence that national representatives
in today's most well-established democracies
are often much more responsive to moneyed interests and lobbyists
than to their electorate~\cite{shlapentokh11feudal,smith14political,cost15republic,flavin15campaign}.

Perhaps a better way to address the coercion risk
is to keep each individual's roles as {\em voter} and potential {\em delegate}
strictly separate in designing an electoral system supporting delegation.
An individual's actions in the voter role would always be private,
and her actions in the delegate role always public and hence accountable.
In her delegate role,
a voter could publicly wield the delegated votes of her followers (if any)
in one way,
but in her voter role privately
cast a vote different from -- perhaps even opposite of -- her public stance,
\ie, secretly {\em declining to support her own public position}.
A potential coercer or vote-buyer can ensure
that she takes the coercer's stance publicly,
but cannot ensure that she actually casts her vote that way privately.
If she doesn't, then her delegate role will wield {\em no} actual voting power
unless other voters independently choose to delegate to
and support her public platform.
Since all of these original votes are likewise private and annonymous,
the would-be-coercer can buy the coercee's public platform
but cannot actually coerce anyone (even the coercee herself) to support it:
if it receives any delegated support at all, it is through the free choice
of individual voters casting their delegation choices anonymously.
The coercee is then ``coerced'' only in the
not-necessarily-harmless but standard and fimilar mode of a celebrity
whose face and participation is purchased in an advertising campaign:
the coercer can do no more than try to make the purchased platform attractive.

Viewed another way, from a ``microparty'' perspective,
we might in principle allow anyone to create a microparty regardless of support,
and to publish a corresponding  ``microparty platform''
with recommended choices on candidates and issues in an election.
This public microparty platform may be vulnerable to coercion or buying,
by virtue of being public as is required for accountability.
However, no one --
{\em not even the ``leader'' of the microparty who registered it} --
can actually be coerced to vote according to the party's public platform.

To ensure that a coercer can't determine
whether a delegate or microparty leader
actually votes for her own public platform,
the electoral system must be designed
not to publish exact support statistics
for delegates or microparties who receive a small amount of support
(\eg, zero, one, or a few votes).
This could be accomplished either by publishing voter support statistics
only for delegates or microparties that achieve some threshold of support,
or by deliberately adding noise to the published statistics
to guarantee the {\em differential privacy} of voters
who do (or don't) support the delegate or microparty~\cite{dwork08differential}.
The support publication threshold or differentially private noise level
might need to be set considerably higher in areas where
not just individual but {\em collective} coercion might be a risk:
\eg, a crime lord threatening a local community with collective punishment
unless most or all of them delegate their vote to him or support his microparty.


\subsection{Delegation to Tools or Algorithms}

\xxx{ cite and discuss \cite{zhang19statement}}

\xxx{
\eg, smart contracts for advice-combining,
allowing voters to ``install'' custom decision-making aids
that combine information from multiple sources in voter-defined ways.

Attractive from a voter choice perspective,
but brings important risks:
especially the danger of voters using tools that they don't understand
that implement algorithms they don't understand
and might be acting against their interests, perhaps maliciously.
(Example: tool that says it's supporting warm fuzzies
but casts votes on behalf of the voter in the opposite direction,
depending on the voter not paying close enough attention.)

Transparency would be critical.
Ensure that any tool/algorithm used is publicly registered,
analyzable by anyone.
Ensure that it's also public knowledge which ones get {\em used} most,
to ensure that the analysis of the public and relevant experts
is explicitly drawn to the ones that could pose the most broad systemic risk
if they behave unexpectedly or are subverted in some way.
}

\xxx{ discuss risks clearly! }

As an alternative to delegating decisions to organizations or individuals,
voters may wish to delegate some of their decisions to,
or ``take advice'' from,
automated tools or algorithms.
In fact this is already occurring in practice,
through voters' growing use of
{\em voting advice applications} or
VAAs~\cite{garzia12voting,louwerse14design,israel17cognitive}.
These tools often take the form of convenient web sites
that ask voters questions about their opinions,
and use algorithms to advise them
about how well- or poorly-matched particular candidates are to their positions.
A 2012 survey found that 20--40\% of the electorate had used VAAs
in recent elections in several European countries~\cite{garzia12voting}.
On the one hand, such applications appear attractive
as tools to empower voters and help them make more well-informed decisions.
On the other hand, such tools present significant risks,
as the algorithms they implement are often non-transparent,
not readily understandable or explainable to most of their users,
and the advice they give can be affected (intentionally or not)
by many subtle design factors.

\com{
``This body of research on VAA methodology can be boiled down to one crucial finding: namely that the design of the tool matters.''~\cite{garzia12voting}

``A small but significant proportion of the tools’ users do declare that they will switch their vote in accordance with the advice obtained.''~\cite{garzia12voting}

``The findings show that irritation emerges if the preferred party is not positioned at the top of the VAA result list. In turn, a strong irritation can lead to a change in vote choice.''~\cite{israel17cognitive}
}

Even among the larger population of voters
not consciously seeking algorithmic advice on their decisions, however,
a high percentage of voters are unquestionably
relying on and indirectly influenced by the algorithms
underlying their social network newsfeeds
to select, filter, learn about, and discuss events,
and to inform their political opinions~\cite{polonski17artificial,landemore18democracy}.
Sensationalistic but false news stories,
produced either by ideologically- or advertising-profit-motivated actors,
can spread more effectively via social media
and potentially influence voters'
opinions~\cite{allcott17social,pennycook18who}.
The fact that these influence-mediating algorithms are typically proprietary,
controlled and understood only by a few engineers in tech giants,
has sparked considerable concern about
algorithmic transparency~\cite{diakopoulos16algorithmic,datta16algorithmic,selbst17meaningful}.
Fake accounts operated by {\em social bots}
can further amplify the spread of misleading or fake
information~\cite{woolley16automating,ferrara16rise,woolley17computational,broniatowski18weaponized,shao18spread}.

\com{	From woolley17computational:
``Ultimately, therefore, we find that bots did affect the flow of information during this particular event. This mixed methods approach shows that bots are not only emerging as a widely-accepted tool of computational propaganda used by campaigners and citizens, but also that bots can influence political processes of global significance.''
}

Given the clear and considerable potential risks inherent
in empowering voters to delegate parts of their thinking to algorithms,
we must proceed with extreme caution in even approaching the idea
of endorsing such practices,
let alone incorporating such capabilities into electoral systems.
Nevertheless, the issue may in the end boil down to one of transparency.
We ultimately cannot prevent users from taking voting advice from algorithms,
either consciously (\eg, through VAAs)
or unconsciously (\eg, through social media newsfeeds),
without also taking away their freedoms of privacy and choice.
Given this impossibility of prohibition,
could we at least design future electronic voting and decision systems
to use more transparent and less risky methods of following algorithmic advice
when they choose to do so?

One way we might envision supporting {\em transparent algorithmic delegation}
is by designing an electoral system supporting advice-giving algorithms
implemented in a fashion similar to smart contracts
in blockchain systems such as Ethereum~\cite{wood14ethereum}.
Such a system might offer users the convenience of being able to use
chosen VAA-like tools available directly within the e-voting system,
like programmable ``plug-ins'' for the voting system.
In exchange for this convenience, however,
the election system would enforce transparency
of both {\em code} and {\em usage}.
By transparency of code we mean that, like Ethereum smart contracts,
in order to operate at all,
the code (software) implementing these VAA plug-ins 
would have to be publicly registered
(perhaps ``on the blockchain'' if the voting system is blockchain-based)
and open to anyone to inspect for potential bugs or hidden malicious behavior
(\eg, algorithmic bias or influence manipulation attempts).
By transparency of usage we mean that
when users {\em do} elect to invoke such plug-in tools,
the electoral system automatically gathers and publishes accurate statistics
on the prevelance of their usage,
ensuring that the public and experts alike can focus
appropriate levels of attention to analyzing
the most popular tools
that could potentially influence significant voting power.

Even if we take the position that voters ``should'' think for themselves,
it may be worth viewing delegation of choice to algorithms
as a human vice not entirely unlike drug abuse or prostitution:
a behvaior that creates important risks --
to democratic health rather than public health --
that may most readily be mitigated
not by either ignoring or banning the practice outright,
but rather by bringing it out of the shadows
where it can be monitored and tightly regulated.

%% file: liquid/spec.tex
\section{Liquidity in Scaling Direct Democracy through Specialization}
\label{sec:liquid:spec}

We have explored above how a liquid notion of voting power
can potentially enhance a voter's choice in answering a given ballot question:
\eg, spreading, transfering, or delegating choice
among candidates running for an office or parliament.
An important complementary issue, however,
is {\em how many and what kinds of questions}
a given election or ballot asks voters to weigh in on.
The ideal of direct democracy is to involve voters {\em directly}
in deciding important policy questions that concern them,
typically through initiatives and referenda.

\subsection{The Ballot and Voter Attention Scalability Problems of Direct Democracy}

Even in countries like Switzerland where direct democracy is deeply rooted,
however,
the extent to which citizens can in practice be directly involved in decisions
is severely constrained by the limited time and attention of the electorate.
Even while holding votes far more often than in most countries
(typically four times per year),
and localizing many direct decisions to cantons or municipalities,
Switzerland must impose fairly high barriers
(\eg, 100,000 signatures for popular initiatives
in a country of only 8 million)
to keep the number of questions on each ballot --
and the magnitude of the task of informing the voters about those questions --
manageable.

\xxx{ cite stats on low participation in Switzerland? }

The critical tension between
the desire to involve voters directly in decisions,
and the need to limit the number of such decisions on the ballot,
stems from one basic premise in traditional designs for direct democracy:
that {\em every voter} needs to be asked, and expected,
to weigh in on {\em every question} in each election.
Constrained by this premise,
direct democracy faces a fundamental scalability limitation.
Since the number and breadth of policy questions affecting a society
is likely to scale in proportion to the size and complexity of that society,
a ``pure'' direct democracy would require a linear $O(n)$ number of voters
each to answer a number of questions at each election
that may similarly grow linearly $O(n)$ with population size,
resulting in an unscalable quadratic $O(n^2)$ overall decision workload.
But could we circumvent this fundamental decision scalability barrier
if we relaxed the presumption that {\em all} voters
should be expected to answer {\em all} questions?
This is a central goal of a number of liquid democracy proposals,
particularly Ford's~\cite{ford02delegative},
which gives voters a ``meta'' level of choice
to answer some questions directly while delegating others to a representative.

\xxx{ direct democracy can fail Dahl's requirement of inclusiveness
by excluding voters due to {\em lack of leisure time}. }

\subsection{Inspiration: Hierarchically Structured Online Discussion Forums such as UseNet}

Concurrent with the early evolution of the Internet,
the first global, decentralized, public electronic discussion forum emerged
in the form of UseNet~\cite{hauben97netizens}.
A precursor to today's popular discussion platforms
like \href{https://slashdot.org}{Slashdot}
and \href{https://www.reddit.com}{Reddit},
UseNet enabled millions of users from thousands of organizations
to post messages in subject-oriented {\em newsgroups}
that anyone elsewhere in the world could read
and respond to in followup debate.
At its height of popularity in the 1990s,
UseNet inspired the publication of
dozens nonfiction guides to UseNet,
studies of its users'
behavior~\cite{okleshen98usenet,kayany98contexts,fiore02observed,turner05picturing},
and even interstellar analogs in science fiction~\cite{vinge93fire}.
Most relevant to our purposes,
UseNet offered a censorship-resistant forum
open to anyone to speak or debate,
but also proved {\em scalable} by virtue of allowing each user
to choose which subset of the roughly 150,000 newsgroups
to read and participate in organized into a deep hierarchy of topics.

UseNet never pretended to offer
a forum for rigorous democratic deliberation,
and the censorship-resistance of its decentralized structure
eventually enabled spammers to overrun it~\cite{cranor98spam},
sending most of its users
scurrying away to more closed or controlled forums.
Nevertheless, many of UseNet's ``netizens''~\cite{hauben97netizens}
were inspired by the perceived ``democratizing'' potential
of a technology platform offering anyone the freedom to speak online
as much or little as they choose, on any topic that interests them --
and to give {\em readers} the necessary corresponding freedom
to manage their limited time and attention by choosing which topics
to follow closely and which to ignore or leave to others.
This perceived democratizing potential of scalable online forums
doubtless inspired many of the variants of liquid democracy
that were proposed in the early 2000s.
This leads to a question that is still as relevant as ever
despite UseNet's failure:
is it possible to enable open, public debate and deliberation,
akin to the processes of direct democracy,
that would scale in richness, participation, and user choice
like UseNet did at its height?
And could such a forum be created that both offers strong freedom of speech
without being overrun by spam, trolling,
and other forms of abuse~\cite{zhai16anonrep,cohen18democracy,landemore18democracy}?

\xxx{ mention UseNet as a prototype of both the promise and peril
of online virtual communities that in principle
support inclusion based on freedom of expression and association
based on shared interests --
which in one sense ensured extremely strong freedom of expression
through the perceived anonymity (``no one knows you're a dog''),
but on the other hand turned hostile and a tool for exclusion
of anyone without 
infinite amounts of time, patience, or tolerance of abuse (trolling etc).
}

\xxx{ clarify the link with ~\cite{cohen18cdemocracy} somewhere }

\subsection{Scalable Direct Democracy via Topic Specialization and Delegation}

Let us now take it as given that we cannot lower the barriers
to voters proposing questions for direct democratic debate and decisions
(\eg, initiatives and referenda),
and thereby vastly expanding the potential number of questions ``on the ballot''
at any given time,
without also giving users a choice of {\em which} (typically small)
subset of those questions to pay attention to, debate, and vote on directly.
We could certainly manage
a large number of topics for direct democratic discussion
by organizing them UseNet-style into a hierrarchy of topics,
giving each user free choice of which to follow and which to ignore.
So let us suppose such a topic hierarchy exists,
and for simplicity assume for now that actual deliberation and choice
occurs only in the bottommost ``leaf'' subtopics
that are not further subdivided.

If only the users with enough time and interest
to follow a particular subtopic closely
actually wield votes on decisions related to that topic, however,
then each small subtopic will behave as a special-interest group,
narrowly representing the interests and opinions
of the specific sub-population of voters
for whom that topic is of prime importance.
For legitimacy, democratic debate and decisions on a topic
somehow need to represent and account for the interests
of the {\em whole} population the topic affects,
not just those few with the time to make that topic their focus.
Healthcare policy affects everyone,
not just doctors and healthcare industry experts;
economic policy affects everyone,
not just bankers and economists.
This need for representativeness remains applicable,
and perhaps becomes even more important,
as we descend in our hypothetical topic hierarchy to narrower subtopics:
\eg, from healthcare as a ``level 1'' topic
to a health insurance subtopic at level 2 underneath it,
from there to a subtopic on pre-existing conditions policy at level 3,
and to a subtopic on pre-existing conditions policy 
specifically for cancer patients at level 4.
A tiny percentage of the total population will likely be able to
follow debates and participate directly in decisions
on such a low-level subtopic,
but nevertheless these decisions affect everyone,
because everyone is exposed to the risk that
they {\em might} get cancer and need insurance coverage if they do.

Representative democracy addresses this legitimate representation challenge
by delegating all these decisions to a handful of elected representatives,
at the cost of effectively overloading that handful of representatives
with so many complex policy decisions that the representatives themselves
can't hope to become experts in {\em any} of them.
These representatives thus focusing instead
on {\em the profession of politics itself} --
being good at campaigning and getting (re-)elected,
rather than being good at making good policy decisions on any one topic.
These professional politicians then necessarily build and rely on
increasingly-vast unelected bureaucracies
to examine and decide policy on
most meaningful questions requiring domain expertise.
This structure necessarily focuses tremendous pressures
on the representatives and the bureaucracies they oversee,
exposing them to strong incentives to corruption through processes such as
professional lobbying and regulatory capture.

Ford's proposal~\cite{ford02delegative}
addresses this tension
between {\em topic specialization} and {\em representation}
by ensuring that {\em all} voters can wield voting on {\em all} topics,
and hence are represented in decisions on those topics.
Since most voters will not have time to follow most of those topics directly,
they instead delegate their vote to a representative,
analogous in principle to the tradition of delegating choices to parties
(Section~\ref{sec:liquid:choice:deleg}),
but with the greater freedom to delegate
to microparties (Section~\ref{sec:liquid:choice:microparties})
or individuals (Section~\ref{sec:liquid:choice:individuals}).
Further, instead of having to choose just {\em one} [micro]party
or individual to represent them generally,
voters could delegate decisions on different topics
to different representatives:
\eg, delegating their vote on healthcare decisions
to a doctor or local medical expert they trust to represent their interests,
and separately delegating their vote on economy decisions
to a local economist or small-business owner whose opinion they trust.
If a voter's immediate delegate cannot follow all aspects (\eg, sub-topics)
of the given topic directly either,
the delegate can transitively {\em re-delegate} decisions on those sub-topics
to other more-specialized experts on those topics.

In a UseNet-like hierarchy of deliberative forums,
each {\em active} participant in a given topic
would thus wield the accumulated power
of all voters who delegated to them directly or indirectly,
as in Alger's proxy voting scheme~\cite{alger06voting}.
Representatives could actually focus on and {\em specialize}
in particular topics
rather than on the generic profession of politics,
accumulating considerable voting power in their topics of specialty
while remaining accountable to and hence legitimately representative of
all the voters who had delegated to them.
Since a representative who is widely-recognized as a specialist in one topic
(\eg, healthcare)
is unlikely to be similarly widely-recognized as a specialist in another
(\eg, economy),
the outsize influence of these specialists on decisions in their focus topic
should not generally translate to outsize voting power on {\em other} topics,
thereby reducing the amount of {\em generic} power any one representative has,
and mitigating both the opportunity and incentive for corrupting influences.

\paragraph{Scalability and Power Spreading from Topic Delegation:}
To illustrate concretely
how topic-specialized delegation could enable direct democracy to scale,
suppose for simplicity that an online liquid democracy forum is organized into
1,000 unique topic-focused forums organized in a hierarchy
consisting of ten top-level level 1 topics (health, economy, etc.),
each of those subdivided into ten level 2 subtopics (health insurance, etc.),
and each of those in turn subdivided
into ten level 3 subtopics (pre-existing conditions, etc.).
Suppose that in a city of 100,000 voters,
the ``average'' voter has time to follow and deliberate directly in
{\em only one} leaf-level subtopic or group in this hierarchy.
Assuming (unrealistically but also unnecessarily)
that the voters' interests are relatively balanced among topics,
this implies that on average about 100 voters will follow and participate in
each level 3 subtopic,
with about 1000 voters following each level 2 subtopic,
and about 10,000 voters following each level 1 topic.
In order to be represented in all of
the 999 level 3 sub-topics they can't follow directly,
each voter needs to make at most nine delegation choices at each level:
who to delegate their vote to in the nine level 1 topics
outside their focus area,
who to delegate their vote to
in each of the nine level 2 sub-topics outside their focus
but {\em within} their level 1 focus area,
and so on.
Voters need not necessarily make each of these nine choicse at each level
separately, either:
providing their delegates have power to {\em re-delegate} transitively,
many voters might simply delegate their voting power outside their focus area
to a ``generic'' delegate they trust to choose specialty delegations for them.

In short, the total amount of decision-making effort demanded of each voter
for ``complete'' participation and representation in this fashion
is logarithmic $O(\log n)$ in the total number of topics $n$,
and hence scalable to large populations and topic hierarchies.
Consider an even-more-ambitious future
in which a country of one {\em billion} voters
implemented such an online liquid democracy forum
with a {\em million} ($10^6$) topics,
subdivided by factors of 10 similar to the example above.
Then if the average voter directly follows and specializes in
only one of the million level 6 topics,
there could be a healthy ``parliament'' of about 1,000 participants in each,
and each voter could still be represented in each of the other 999,999 topics
by making at most $6 \times 9$ delegation decisions.
In practice, voters seem likely to simplify their choice further
by delegating to the same semi-generic friend or colleague
in several of the sub-topics at a given level,
perhaps making only {\em one} generic delegation choice at one or more levels,
thereby delegating to their chosen generic representative
the task of deciding which specialists to delegate to
on particular subtopics.
Regardless of how fine- or coarse-grained a voter's delegation choices are,
they are also likely to keep many of these delegation choices ``persistent''
across time and direct democracy election events,
rather than re-evaluating them before each new decision,
amortizing the decision costs of making even fairly complex delegation choices.

Because each of these ``average'' voters may accumulate
considerable delegated voting power in their specialty
but is unlikely to accumulate much delegated power far from their specialty,
the incentive and vulnerability to corruption among these specialized delegates
may be many orders of magnitude less than in a comparable society in which
(say) a {\em single} traditional parliament of 1,000 members
generically represented a million users each on {\em all} policy matters
in this billion-voter country.
While each representative of the generic parliament would wield
one million times the political power of an ordinary individual
on {\em all matters} -- a tremendous power concentration --
a specialized representative in each of the million liquid democracy topics
would wield a million times the voting power of a single individual
but only on about {\em one millionth} of the issues being debated and decided,
exercising little to no delegated power in other specialties.
Even accounting for the fact that in practice we can expect
balances of delegated voting power to be considerably less even
(\eg, outsize amounts of generic power perhaps going to some celebrities),
nevertheless the spread of power across a multitude of specialized topics
seems likely to result in far less concentration of power in general,
and hence perhaps susceptibility to corruption,
than in a traditional representative government at large scale.

\subsection{Scalable Choice at Varying Levels of Specialization}

We assumed above that online deliberation and democratic decision-making
occur only at the bottommost leaf levels of the topic hierarchy,
but this constraint is probably neither strictly necessary nor desirable.
Within each low-level subtopic,
issues are likely to arise that appear particularly important or contentious,
which may merit pushing those issues ``up the hierarchy''
to less-specialized subtopics or topics that more voters follow.
For example, a contentious policy decision
on pre-existing conditions for cancer patients
might start in the corresponding low-level subtopic
but then ``bubble upwards'' to become an initiative or referendum
presented directly to, and inviting debate and decision participation from,
voters in higher-level topics such as insurance policy
or healthcare policy in general.

Since issues that bubble up the topic hierarchy this way
demand the attention, enlightened understanding, and decision capacity
of progressively larger numbers of voters
($10 \times$ per level in our hypothetical examples above),
we must impose barriers or thresholds on this upward propagation
to ensure that each voter needs to examine and decide on
only a manageable few issues at each hierarchy level in each time period.
A few such issues each cycle might capture particularly broad interest,
surmounting the highest thresholds,
and as a result propgating all the way up to the topmost level,
where such issues are presented to the {\em entire} voting population
for consideration and decision.
Issues that propagate all the way to the top in this way
are thus precisely analogous to popular initiatives
in today's working direct democracies such as Switzerland,
where only a few issues passing high signature thresholds
go onto the ballot that all voters see.
Some issues are indeed important or of broad enough interest
that it is worth presenting them to the whole electorate --
but since the number of these necessarily must be limited,
a topic hierarchy with liquid delegation of choice
could enable a much larger number of direct democratic deliberations to occur
while preserving representativeness at all levels
and allowing each issue to ``bubble up'' to the level of specialization
corresponding to the breadth of voting population that considers the issue
worthy and important enough for consideration at that level.

\subsection{Choice and Management of the Topic Hierarchy}

\xxx{ cite Dahl's principle that voters need ``control of the agenda'' }

The above discussion has assumed
that a suitable topic hierarchy is somehow given,
but the organization of this hierarchy is of course
another topic that would have to be decided somehow.
A simple approach would be to delegate the task of tending this hierarchy
to representatives elected at large, \eg, a traditional parliament.
While feasible, this approach would present the risk
that representatives interested in their own re-election
might be strongly tempted to structure the topic hierarchy
in favor of their interests,
in an analog of the real present-day problem of gerrymandering.
Does a low-level subtopic related to the link between air quality and cancer
belong under the topic of Healthcare or Environment?

This choice matters,
because the location of the subtopic in the hierarchy
will affect how voting power gets delegated and hence distributed to
direct participants in that subtopic.
An elected representative with strong ties to the healthcare industry
might benefit from tweaking the hierarchy
so that his healthcare-specialist friends get
as much delegated voting power as possible in decisions on the subtopic,
while a representative with stronger ties to environmentalist groups
might have the opposite preference
and fight to re-locate the subtopic under Environment.
While contention around such hierarchy-management issues
is probably unavoidable given
the human judgment and subjective preferences ultimately involved,
it generally seems preferable if possible
to avoid hierarchy management becoming
yet another political football -- like districting choicse -- 
to incentivize and attract corruptive tendencies.

A safer alternative, therefore,
might for the {\em electorate at large} to decide the manage hierarchy directly,
using the same proportional-representation tools
we already commonly use for eleecting parliaments for example.
Suppose we stipulate by design that there shall be
at most ten top-level topics at level 1,
ten level 2 subtopics under each level 1 topic, etc.
We might allow anyone to propose keywords they think should be level 1 topics,
and periodically use a multi-winner proportional-representation system
such as STV, CTV, or QTV (Section~\ref{sec:liquid:choice:trans})
to ``elect'' proposed keywords (instead of human candidates)
into the 10 available ``positions'' of level 1 topic status.
The election system's proportionality measures ensure
that the ten chosen level 1 topics represent those
of broadest interest to the electorate at large,
while ensuring that each voter's choice counts
and as few votes as possible are ``wasted.''
Once level 1 topics have been chosen this way,
a similar multi-winner proportional election may be held
within each of these level 1 topics to its level 2 subtopics, and so on.
Since not every voter will likely be even aware of,
let alone interested in or knowledgable about,
all of the detailed topics at the lower levels of the hierarchy,
enabling users to delegate their power to specialists at these lower levels
will be just as important to setting up the lower levels of the hierarchhy
as to facilitating representative decisions within them subsequently.

No single topic hierarchy, once chosen, will remain ``perfect'' forever,
of course,
so processes for revising and evolving the hierarchy will be just as essential.
One simple approach is again to treat elections of keywords to topic status
much like electing candidates to parliamentary seats:
namely, ensure that each keyword's status
goes ``up for re-election'' periodically
(\eg, once every few years),
giving the electorate the opportunity to ``promote'' topics to higher levels
and to ``demote'' topics whose interest has shrunk to lower,
more specialized levels underneath other broader topics.
A change in topic status at a high level may need to trigger
an automatic ``re-election'' at lower levels containing related subtopics,
to ensure that deliberation and decisions
that used to occur in a newly-demoted topic
have a suitable place to move to in the lower levels of the hierarchy,
for example.

\subsection{Expressing Strength of Interest: Quadratic Voting in Participation Scalability}

\xxx{ note that quadratic voting doesn't solve this problem...
	strawman: flat.  distraction politics. 
	hierarchical alternative.}

Given that the intended purpose of quadratic voting (QV)
is to enable and incentivize people to express their true stength of preference
in decision-making~\cite{posner15voting,lalley18quadratic},
an obvious question is whether QV
could address the scalability challenge to expanding direct democracy.
Suppose, for example, we lower the traditional barriers
to getting popular initiatives and referenda on the ballot,
allowing perhaps hundreds or thousands of questions onto the ballot each cycle
rather than just a few.
Can we then just give each voter an equal amount of virtual ``coin''
with which to buy votes on the particular issues they care about,
as the QV theory suggests?

Deploying QV this way might be attractive from an economic purist's perspective,
but could be profoundly dangerous and counterproductive in practice,
due to the gap between the ``perfectly rational voter'' QV theory assumes
and the imperfect understanding of human voters.
As discussed earlier in Section~\ref{sec:liquid:choice:spread:quad:distract},
a realistic voter who is unaware
how much a moderately-subtle policy issue affects him
will ``underspend'' on that issue,
perhaps investing nothing at all to vote either for or against
many such issues that he is not adequately informed about.
Furthermore,
the practice of distraction politics~\cite{jamieson93dirty,weiskel05sidekick,leibovich15politics}
could deliberately draw voting power away from meaningful but subtle issues
toward ``bright shiny objects'' and sensationalistic controversies,
effectively handing most decision power on most special-interests 
to the special interests that care about them.

Quadratic voting therefore does {\em not} by itself
address the scalability problem that delegation in liquid democracy addresses,
although the two could potentially be used in complementary ways.
One obvious approach is to use QTV
(Section~\ref{sec:liquid:choice:trans:quad})
to enable voters in choosing and maintaining the topic hierarchy,
as discussed above.
In this case,
we use quadratic voting only to choose the
(\eg, ten) topics of broadest interest to form the first-level topics,
and similarly using separate QTV instances within each topic
to choose up to ten sub-topics to flesh out the hierarchy.
It is much more reasonable to assume average voters are well-equipped
to make reasonable judgments on the {\em ten} broad policy topics
of most importance to them,
than to make reasonable judgments on the importance of --
let alone actual decisions on --
hundreds or thousands of narrow specialized topics.
Provided delegation is permitted and in effect 
in the QTV instances that decide on sub-topics of an already-specialized topic,
we have at least some hope that the bulk of voting power
used to decide those subtopics
is by then (through delegation) in the hands of people
who are both specialized enough to follow the topic in question,
and representative of and accountable to those whose voting power they wield.

\xxx{ 	maybe too niggly and subtle...
Another variation of this approach worth considering
is to use QTV within each topic in the hierarchy as above,
but to eliminate the election quota and vote transfers from elected candidates
from QTV,
and instead let QTV produce a selection of the top ten (sub)topics 
while allowing them to have {\em uneven} voting weights.
}

Another way it might be useful and arguably ``safe''
to use quadratic voting in this context
is to allow users to assign different levels of importance
among a {\em small} number of different ``peer'' topics or subtopics
relative to each other in the hierarchy,
and adjust the user's ``voting weight'' in each of those subtopics accordingly.
Suppose for example that the ten top-level topics
include Healthcare and Environment,
and Alice cares much more strongly about healthcare-related policy
than about environmental policy.
Instead of automatically giving her equal voting power in both topics,
as we normally would expect by default,
we might give her the option to divide her voting power by parts
and allow her to assign it either evenly or unevenly to the topics using QV
according to her judgment of their relative importance to her.
For example, if we considers Healthcare far more important than Environment,
she might assign four parts of her voting power to the former
and only one to the latter,
and thereby wield a vote on Healthcare issues having $2\times$ the power
as her vote on Environment issues.
Within a particular lower-level topic
she follows and is particularly interested in
(\eg, health insurance policy),
she might similarly have an option to redistribute her voting power
among the subtopics
(\eg, more to pre-existing conditions policy
and less to pediatrics insurance policy).

The critical point is that in this approach,
she would be using quadratic voting to redistribute her voting power
only among a {\em small} number of closely-related subtopics,
in {\em one} small local area of the hierarchy,
where we can plausibly expect the voter to have some genuine understanding
of those subtopics and their ``true'' importance to her
relative to each other,
givern that she is taking the trouble to perform this redistribution at all.
Thus, {\em local} redistribution of voting weight using QV
among a few alternatives we have reason to expect the user to know about
may plausibly safe and valuable,
whereas {\em global} redistribution of voting weight
among a large number of options mostly unfamiliar to the voter
would seem extremely dangerous and fraught with risk.

%% file: liquid/time.tex
\section{Liquid Democracy and Time}
\label{sec:liquid:choice:time}

{\em In preparation.}

%% file: liquid/impl.tex
\section{Considerations for Systems Implementing Liquid Democratic Choice}
\label{sec:liquid:choice:impl}

{\em In preparation.}

%% file: liquid/risk.tex
\section{Potential Risks}

{\em To be written}

\xxx{

the superstar delegate problem...

importance of vote-splitting...

note interesting idea: to the extent this is a problem,
one potential way to counter it would be to allow vote splitting
{\em and} use quadratic voting to determine the ``return'' on split investment.
This would in effect reward and incentize broader spreading of voting power.
A voter who considers the positions of several microparties compatible
can split his vote among them and help all of them,
and will help them all more in (additive) aggretate than the voter
could help any one of them individually.
A ``plumping'' voter who invests {\em all}
of his voting power in one celebrity-centric microparty
gets less total influence in return for his investment.
\xxx{ discuss earlier how this principle might apply
	in the context of vote-splitting for proportional representation,
	for example, to reward and empower centrist voters 
	who support multiple parties over those
	with strict allegiance to just one. }

Increasing ``under-the-hood'' complexity...
e.g., fractions of votes or ``microvotes'';
graph algorithms and flows...

However, these are the types of complexity increases
that are easy to make transparent and understandable to
{\em anyone who wants to dig just a bit deeper}
than the average voter.
For example, while a substantial fraction of the world's population
(a billion?)
might have at least heard of Bitcoin by now,
a much smaller but still large number of people
(a million?)
might know enough about it to have heard of ``satoshis,''
and perhaps to be distantaly aware that one bitcoin consists of 
a large fixed number of satoshis
(this number happens to be XXX but far fewer people probably know that).
Bitcoin is ``transparent'' in that
anyone with enough technical knowledge can look at the protocol specifications
and the relevant software source code,
examine the details,
and ensure independently that everything does indeed operate as claimed.
Individuals can ask their friends for technical opinions,
and companies can hire experts to analyze risks
or particular software implementations.
The degree of effective transparency varies vastly from person to person,
but the point is that the technology gives
everyone the {\em choice} of to what depth to exercise their {\em option}
to examine all the technical details of the technlogy.

There are other challenges with e-voting technologies in general
that are much more technically difficult to address
than public structural calculations of the kind needed
for liquidity mechanisms like those discussed above.
In particular, ensuring that voters can keep their votes private,
while also enabling them to verify in a public database
that their private vote was correctly recorded and included in the tally,
is a difficult challenge that has nevertheless been mostly solved
by appropriate use of cryptographic techniques such as
verifiable shuffles \xxx{cite} and homomorphic encryption \xxx{cite}.
The challenge of ensuring that voters using an e-voting system
cannot easily be {\em coerced} or their votes secretly {\em bought},
without compromising the system's other privacy or integrity properties,
represents an even harder technical challenge
that is still essentially unsolved,
and is one key reason many security/privacy experts
are generally against using e-voting technologies at all
except perhaps for ``low-stakes'' elections \xxx{cite}.

}

%% file: liquid/rel.tex
\section{Related Work}

{\em To be written}

%% file: liquid/conc.tex
\section{Conclusion}

{\em In preparation.}